\documentclass[twocolumn]{aastex62}
\usepackage[T1]{fontenc}
\usepackage{graphicx}	
\usepackage{amsmath}	
\usepackage{amssymb}	
\usepackage{natbib}
\newcommand{\angstrom}{\textup{\AA}}
\newcommand{\mbh}{$M_{\mathrm{BH}}$}
\newcommand{\dotm}{$\dot{m}$}
\newcommand{\as}{$a_{\ast}$}
\newcommand{\incl}{$i$}
\newcommand{\mbhlit}{$M_{\mathrm{BH}}^{\mathrm{lit}}$}
\newcommand{\mbhbay}{$M_{\mathrm{BH}}^{\mathrm{Bay}}$}
\newcommand{\dotmlit}{$\dot{m}_{\mathrm{lit}}$}
\newcommand{\mbhobs}{$M_{\mathrm{BH}}^{\mathrm{obs}}$}
\newcommand{\dotmobs}{$\dot{m}_{\mathrm{obs}}$}
\newcommand{\Hb}{H{\small $\beta$}}
\newcommand{\MgII}{\ion{Mg}{2}}
\newcommand{\CIV}{\ion{C}{4}}
\newcommand{\FeII}{\ion{Fe}{2}}
\newcommand{\NV}{\ion{N}{5}}

\newcommand{\mmbh}{M_{\mathrm{BH}}}
\newcommand{\mdotm}{\dot m}
\newcommand{\mas}{a_{\ast}}
\newcommand{\mmbhlit}{M_{\mathrm{BH}}^{\mathrm{lit}}}
\newcommand{\mdotmlit}{\dot{m}_{\mathrm{lit}}}

\newcommand{\Ks}{K$_{\rm S}$}
\newcommand{\Msun}{${\rm M_{\odot}}$}
\newcommand{\mMsun}{{\rm M_{\odot}}}
\newcommand{\FWHM}{\textit{FWHM}}
\newcommand{\mFWHM}{\mathit{FWHM}}
\newcommand{\EW}{\textit{EW}}
\newcommand{\mRblr}{R_{\mathrm{BLR}}}

\newcommand{\ergs}{erg s$^{-1}$}
\newcommand{\kms}{km s$^{-1}$}
\newcommand{\mergs}{\mathrm{erg\, s}^{-1}}
\newcommand{\mkms}{\mathrm{km\, s}^{-1}}

\graphicspath{{./}{figures/}}

\begin{document}
\title{Black hole masses of weak emission line quasars based on continuum fit method}

\correspondingauthor{Marcin Marculewicz}
\email{marcin.marculewicz@gmail.com}

\author{Marcin Marculewicz}
\affiliation{University of Bia\l{}ystok, Faculty of Physics, Cio\l{}kowskiego 1L Street, 15-245 Bia\l{}ystok, Poland}
\author{Marek Nikolajuk}
\affiliation{University of Bia\l{}ystok, Faculty of Physics, Cio\l{}kowskiego 1L Street, 15-245 Bia\l{}ystok, Poland}

\begin{abstract}
We studied optical-ultraviolet spectral energy distribution of
10 weak emission-line quasars (WLQs) which lie at redshifts z = 0.19
and 1.43 < z < 3.48. The theoretical models of their accretion disk
continua are created based on the Novikov-Thorne equations. It allows
us to estimate masses of their supermassive black holes (\mbh) and
accretion rates. We determined the virial factor for WLQs and note its
anti-correlation with the full width at half maximum (\FWHM) of
\Hb\ emission-line ($f \propto \mFWHM^{\alpha}, \alpha = -1.34 \pm
0.37$). By comparison with the previously estimated BH masses, the
underestimation of \mbh\ is noticed with a mean factor 4-5 which depends
on the measured full width. We proposed the new formula to estimate
\mbh\ in WLQs based on their observed \FWHM(\Hb)\ and luminosities at
5100\AA. In our opinion, WLQs are also normal quasars visible in a reactivation stage.
\end{abstract}



\section{Introduction}
\label{sec:1}

Weak emission-line quasars (WLQs) are unsolved puzzle in the model of
active galactic nuclei (AGN).  Typical equivalent width (\EW) of
\CIV\ emission line is extremely weak ($\la 10{\angstrom}$) compared to
normal quasars and very weak or absent in Ly$\alpha$ emission
\citep{Fan99,DS09}. \citet{DS09} concluded that WLQs have optical
continuum properties similar to normal quasars, although Ly$\alpha$+\NV\ 
line luminosities are significantly weaker, by a factor of 4. An
explanation for the weak or absent emission lines has not be found
so far. Probable explanations include a radiatively inefficient
accretion flow \citep{Yuan04} and a cold accretion disk
\citep{Laor11} with a small accretion rate.  An extremely high
accretion rate where we have inefficient photoionized flux is proposed
by \citet{Leighly07a,Leighly07b}. Parallel to that
\citet{Wu11} postulate presence of shielding gas between the accretion
disk and a Broad Line Region (BLR), which could absorb the high-energy
ionizing photons from the accretion disk.  The last but not least,
model suggests an unusual BLR i.e., anemic in construction and gas
abundance \citep{Shemmer10, Nikolajuk12}. This explanation supports
the idea that WLQs can also be in the early stage of AGN evolution
\citep{Hryniewicz10, Liu11, Banados14, Meusinger14}.  Several works
about X-ray properties of WLQs have recently appeared
\citep[e.g.][]{Wu11,Wu12,Ni18, Marlar18}.  
Conclusion arising from these works is that WLQs are more likely X-ray weaker (about half of them) than normal quasars. For example, \citet{Ni18} mentioned that 7 of the 16 WLQs in their sample are X-ray weak. \citet{Luo15} suggested that it may be caused from the shielding gas which prevents the observer to see central X-ray emitting region.

Knowledge about the values of black hole (BH) masses and the accretion rates is crucial in understanding an accretion flows phenomena. The most robust technique is the reverberation mapping method \citep[RM,][]{Blandford82, Peterson93, Peterson14, Fausnaugh17, Bentz18, Shen19}. The method is based on the study of the dynamics surrounding the black hole gas. In this way, we are able to determine
the supermassive black hole (SMBH) mass:
\begin{equation}
    \mmbh = 
    \frac{v^2_{\mathrm{BLR}} R_{\mathrm{BLR}}}{G} =
    f \frac{\mathrm{\FWHM}^2 R_{\mathrm{BLR}}}{G}
\label{eq:blr}    
\end{equation}
where \mbh\ is the black hole mass, G - the gravitational constant,
$R_{\mathrm{BLR}}$ is a distance between the SMBH and a cloud in the
broad line region (BLR). $v_{\mathrm{BLR}}$ is a velocity of the cloud
inside the BLR.  This speed is unknown and we express our lack of
knowledge in the form of the Full Width Half Maximum (\FWHM) of an
emission-line and $f$ - the virial factor, which describes a
distribution of the BLR clouds. In the RM the $\mRblr$\ is determined
as the time delay between the continuum change and the BLR
response. This technique requires a significant number of
observations.  The modification of this method is the
single-epoch virial BH mass estimator \citep[see][for review]{Shen13}.  The correlation between $\mRblr$ and continuum luminosity ($\nu L_\nu$) is observed \citep{Kaspi05, Bentz09} and incorporated in Eq.~(\ref{eq:blr}).  Thus, the method is powerful
and eagerly used because of its simplicity \citep{Kaspi2000, Peterson04, Vestergaard06, Plotkin15}.
The non-dynamic method, which is the spectra disk-fitting method, is
based on well grounded model of emission from an accretion disk (AD)
surrounded black hole \citep[e.g.][]{SS73, NT73}.  The most important
parameter in such models is the mass of black hole and the accretion
rate. The spin of the black hole and the viewing angle are also taken
into account.  In this technique, a SED of an AGN is fitted to the
model and one is able to constrain these four parameters.  More
advanced disk spectra models, which take into account an irradiation
effect, limb-darkening/brightening effects, the departure from a
blackbody due to radiative transfer in the disk atmosphere, the
ray-tracing method to incorporate general relativity effects in light
propagation, can be fitted \citep[e.g.][]{Hubeny2000, Loska04, Sadowski09, Czerny11HNS, Laor11,Czerny19BLRscaling}. Generally, the RM and the single-epoch virial BH mass method are
inadequate for BH mass estimation in WLQs, due to weakness of emission-lines 
in these objects.

This paper is organized as follows. Section \ref{sec:2} consists of the description of data selection and reduction. Section \ref{sec:3} explains procedures used to fit models of accretion disk to observations. Section \ref{sec:4} presents our results. Discussions and conclusion are presented in Sections \ref{sec:5} and \ref{sec:6}.
In this work we compute luminosity distances using the standard cosmological model ($H_{0}$ = 70 \kms\ Mpc$^{-1}$, $\Omega_{\Lambda}$ = 0.7, and $\Omega_{\rm M}$ =
0.3 \citep{Spergel}.
\section{Sample selection and data preparation}
\label{sec:2}

\subsection{Sample selection}
\label{sec:2.1}

The sample contains 10 WLQs, which positions cover a wide range of
redshift from 0.2 to 3.5 (see Tab.~\ref{tab:coordinates}). Four
objects, namely SDSS J083650.86+142539.0 (thereafter J0836), SDSS
J141141.96+140233.9 (J1411), SDSS J141730.92+073320.7 (J1417), and
SDSS \linebreak J144741.76-020339.1 (J1447) were analysed by
\citet{Shen11,Plotkin15}. Three next sources -- SDSS
J114153.34+021924.3 (J1141) and SDSS J123743.08+630144.9 (J1237) were
studied by \citet{DS09}, and SDSS \linebreak J094533.98+100950.1
(J0945) by \citet{Hryniewicz10}. The quasar SDSS J152156.48+520238.5
(J1521) was inspected by \citet{Just07, Wu11}.  The number of objects
in the sample from the SDSS campaign \citep{SDSSDR7} has been
increased by two next WLQs: PG 1407+265 and PHL 1811.  The first object is the first observed WLQ in history and intensively examined
by \citet{Mcdowell95}.  PHL 1811 is the low redshift source classified
also as NLS1 galaxy \citep{Leighly07a, Leighly07b}.
\subsection{Observed data}
\label{sec:2.2}

Photometric points of WLQs at visible wavelengths are collected based
on the Sloan Digital Sky Survey (SDSS) optical catalog Data Release
7. It contains the u, g, r, i, and z photometry \citep{SDSSDR7}. In the
case of PHL 1811, we based on measurements of fluxes in B and R
colors, and performed by the Dupont 100" telescope at Las Campanas
Observatory (LCO) \citep{Prochaska11}.  The flux at U band was
observed by the UVOT telescope on-board the {\it Swift} satellite
\citep{UVOT}.  Near-infrared photometry in the W1-W4 bands are taken
from the Wide-field Infrared Survey Explorer (WISE) Preliminary Data
Release \citep{wise,Wise_2}.  Those data were supplied by photometry
in the J, H, \Ks\ colors obtained from the Extended Source Catalog of
the Two Micron All Sky Survey (2MASS) \citep{2mass}.  Crucial points
for the project are those detected in near- and far-ultraviolet (NUV,
FUV, respectively) wavelengths.  They are provided by the Galex
Catalog Data Release 6 \citep{galex}.  

Additionally, we use the spectra observed by SDSS, to check photometric data positions in regards to the spectrum. In the case of PHL 1811 and PG 1407+265 the spectra are taken from \citet{Leighly07a} and \citet{Mcdowell95}, respectively. A basic observational properties of the WLQs sample and sources of their photometry points are listed in Tab.~\ref{tab:coordinates}.

\begin{deluxetable*}{lccccl}
\tablenum{1}
\tablecaption{Sample of Weak Emission-Line Quasars and the sources of their photometry points \label{tab:1}}
\tablehead{
\colhead{Name} & \colhead{RA} & \colhead{Dec} & \colhead{$z_{spec}$} & \colhead{$A_{V}$} & \colhead{Photometry data} \\
\colhead{} & \colhead{(deg.)} & \colhead{(deg.)} & \colhead{} & \colhead{(mag.)} & \colhead{}}
\decimalcolnumbers
\startdata
SDSS J083650.86+142539.0 & 129.211935 & +14.427527 & 1.749 & 0.129& WISE (W1,W2,W3), SDSS (u,g,r,i,z), Galex (NUV) \\
SDSS J094533.98+100950.1 & 146.391610 & +10.163912 & 1.683 & 0.062& 2MASS (J,H,\Ks), SDSS (u,g,r,i,z), Galex (NUV,FUV) \\
SDSS J114153.34+021924.3 & 175.472251 & +02.323508 & 3.55 & 0.065 & WISE (W1,W2,W3,W4), SDSS (u,g,r,i,z) \\
SDSS J123743.08+630144.9 & 189.429435 & +63.029141 & 3.49 & 0.032 & WISE (W1,W2,W3,W4), SDSS (u,g,r,i,z) \\
SDSS J141141.96+140233.9 & 212.924908 & +14.042742 & 1.754 & 0.064 & WISE (W1,W2,W3,W4), SDSS (u,g,r,i,z), Galex (NUV,FUV) \\
SDSS J141730.92+073320.7 & 214.378855 & +07.555744 & 1.716 & 0.084 &  WISE (W1,W2,W3,W4), SDSS (u,g,r,i,z), Galex (NUV,FUV) \\
SDSS J144741.76-020339.1 &  221.924048 & -02.060986 & 1.430 & 0.163 & WISE (W1,W2), SDSS (u,g,r,i,z), Galex (NUV,FUV)) \\
SDSS J152156.48+520238.5 & 230.485324 & +52.044062 & 2.238 & 0.052 &  WISE (W1,W2,W3,W4), 2MASS (J,H,\Ks), SDSS (u,g,r,i,z) \\
PHL 1811 & 328.756274 & -09.373407 & 0.192 & 0.133 & WISE (W1,W2,W3,W4), 2MASS (J,H,\Ks), LCO (B,R), Swift (U), \\
 & & & & & Galex (NUV,FUV) \\
PG1407+265 & 212.349634 & +26.305865 & 0.940 & 0.043 &WISE (W1,W2,W3), 2MASS (J,H,\Ks), SDSS (u,g,r,i,z) \\
\label{tab:coordinates}
\enddata
\tablecomments{
The coordinates (Col. 2 and 3), spectral redshift (Col. 4), and foreground Galactic extinction measured at the V color (Col. 5) are taken from NED. The column (6) contains references to the names of the relevant catalogs and photometric points.}
\end{deluxetable*}

To check if our disk fitting method works with respect to WLQs correctly, we are running a sample method of normal type 1 quasars. For this purpose, we select the sample of objects taken from the Large Bright Quasar Survey (LBQS) \citep{Hewett95, Hewett01}. It is one of the largest published spectroscopic surveys of optically selected quasars at bright apparent magnitudes. It contains data, including positions and spectra of 1067 quasars. Additionally, \citet{Vestergaard09} give black hole masses and Eddington accretion rates estimates of 978 LBQS (see their Table 2). The disk fitting method gives results that we can trust as long as the bend point in SED and the spectrum in the ultraviolet are visible. For this reason, we have chosen 27 quasars with the presence of a well visible big blue bump. The sample of the normal quasars are observed at redshifts between 0.254 and 3.36. Their supermassive black holes masses are in the range 8.09--10.18 [in $\log M_{\rm BH}$ (M$_{\odot}$), Fig. \ref{fig:lbqs}], and luminosities, $\log L_{\rm bol}$ (\ergs) =  45.25--47.89. 
Photometric points of selected quasars come from the same catalogs mentioned earlier.
\subsection{Dereddening} 
The observational data requires corrections, because they are contaminated either by internal or external effects such as a dust in our Galaxy, an influence of the intergalactic medium, starlight from the host galaxy, a dusty torus in the AGN. Firstly, the Spectral Energy Distribution (SED) of all objects are corrected for Galactic reddening with an extinction law. This extinction curve is usually parameterized in terms of the V-band extinction, $A_{\rm V}$, and a measure of the relative extinction between B and V-band: $R_{\rm V} = A_{\rm V}/E(B-V)$. The value of $R_{\rm V}$ varies from 2.6 to 5.5 in the measurements of the diffuse interstellar medium with a mean value of 3.1 \citep{Cardelli89,Fitzpatrick99}.  $A_{V}$ values are taken from NED\footnote{The NASA/IPAC Extragalactic Database (NED): {\tt\string ned.ipac.caltech.edu} } based on the dust map created by \citet{Schlegel98}. \citet{Cardelli89} extinction curve has cutoff at 1250 \AA\ and some photometric points we use go back to shorter wavelengths. Nevertheless, the extinction law examined in the range of 900-1200 \AA\ seems to follow the Cardelli et al. law \citep{Hutchings01}. In this way, we extrapolate the curve down to 900\AA\ for our FUV photometric points by using the same formula.
\subsection{UV and SDSS photometric points' correction}

In the case of high-z quasars, the UV fluxes are very sensitive to
photoelectric absorption in the intergalactic medium (IGM).  We do not
know an attenuation of the flux by the IGM along each line of sight. 
Therefore, following \citet{Castignani13}, we have used the effective optical
depth $\tau_{\rm eff}(\nu, z)$, which are averaged over all possible
directions.  Based on the values collected in Castignani's et al. Table 1,
we correct the observed intensity: $I_{\rm
  \nu, em} = I_{\rm \nu, obs} \exp(\tau_{\rm eff}(\nu, z))$. 
 Thus, we recalculate the fluxes in the Galex FUV
and NUV, the Swift U band, and the SDSS u, g filters. The
effective optical depth in the three other SDSS filters (i.e. r, i,
and z) vanishes for the redshift range considered here.

\subsection{Starlight}
\label{contamination}

A contribution to the SED from stars in the QSO host galaxy is likely
to be negligible \citep{Shen11_starlight,Collinson15}.  Nevertheless,
we would like to check its contribution and we determine a level of
the starlight for each of objects individually.  Following
\citeauthor{Collinson15}, we use a 5 Gyr-old elliptical galaxy
template\footnote{SWIRE Template Library: \citet{Polletta07};
{\tt\string www.iasf-milano.inaf.it/$\sim$polletta/templates/}} as the
stars contamination to the fluxes. We estimate the level of the starlight using the
\mbh--$L_{\rm bulge}$ relation \citep{Degraf15}, where $L_{\rm bulge}$
is the bulge luminosity in the V-band (cyan line, in
Fig.~\ref{fig:SUM}). \mbh\ are the BH masses in WLQs collected from
literature (see Col.~7 in Tab.~\ref{tab:bestfit}).
We noticed that starlight has bigger contribution to WISE points than
to the optical/UV data and adding contamination of the starlight to
those data helps us refine the fitting procedure.

\subsection{Torus contamination}

Together with accretion disk emission, we fit one or two
single-temperature blackbody (BB) as a thermal emission of tori
visible at IR data (see Fig.~\ref{fig:SUM}). The fitted temperatures are collected in 
Tab.~\ref{tab:Temperature}. The mean values of them for
the two BB components are 1110 K and 460 K, respectively. Those values
are close to those referred as 'hot' and 'warm' BB components (1100-2200K
vs. 300-700K) by \citet{Collinson17}.  The temperatures of 'hot'
component are also similar to those seen in WLQs 
($\mathrm{870\,K < T < 1240\,K}$; \citealt{DS09}).

\begin{table}
   \tablenum{2}
    \centering
    \caption {Fitted temperature of torus}
    \begin{tabular}{cc}
    \hline   
    Name & Temperature [K]\\
    \hline
    J0836 & 450, 1400 \\
    J0945 & - \\
    J1141 & 450, 970 \\
    J1237 & 430, 970  \\
    J1411 & 410, 970  \\
    J1417 & 430, 970  \\
    J1447 & 410, 1250  \\
    J1521 & 750 \\
    PHL 1811 & 450, 1100  \\
    PG 1407 & 650, 1240  \\
    \hline
    \label{tab:Temperature}
    \end{tabular}
 \end{table}

\section{Method}
\label{sec:3}
\subsection{Model of an accretion disk}

The primary goal of this work is to fit SED of quasars by the simple
geometrically thin and optically thick accretion disk (AD) model
described by \citeauthor{NT73} (NT) equations.  In the simplest
approach, the AD continuum can be illustrated by the \citeauthor{SS73}
model, nevertheless this attitude does not include a non-zero
spin. The solution to this problem has resulted in the NT equations
that we use in our numerical code. As the spin of the black hole
increases, the innermost stable circular orbit (ISCO) decreases and
the disk produces more high-energy radiation. The output continuum of
the NT model is fully specified by four parameters, which we
determine. These 4 parameters are: the black hole mass -- \mbh, the
mass accretion rate -- $\dot M$, the dimensionless spin\footnote{$\mas
  = \frac{cJ}{GM^2}$} -- \as, and the inclination -- \incl\ at which
an observer looks at the AD. The mass of the black hole is expressed
in units of mass of the Sun (\Msun), and the accretion rate in the
form of the Eddington rate, i.e. $\dot m = \dot M/\dot M_{\rm Edd}
\propto \dot M/\mmbh$.  We construct a grid of 366000 models of AD,
for evenly spaced values of \mbh, \dotm, \as, and \incl.  The $\log
\mmbh$ range is from 6.0 to 12.0, the Eddington accretion rate covers
the band 0--1, and the dimensionless spin $0 \leq \mas \leq 0.9$ with
the step 0.1.  The inclination is fixed for 6 values that cover a
range from 0$^{\circ}$ to 75$^{\circ}$ with the step of 15$^{\circ}$
(see Tab. \ref{tab:grid}).
\tablenum{3}
\begin{table}
\centering
\caption {Parameter values for the grid of the AD models \label{tab:2}}
\begin{tabular}{lcc}
\hline
\hline
Parameter & $\Delta$ & min-max values\\
\hline
 $\log$ \mbh  & 0.1 & 6--12\\ 
 \dotm & 0.01 & 0--1 \\
 \as & 0.1 & 0--0.9 \\
 \incl & 15$^{\circ}$ & 0$^{\circ}$--75$^{\circ}$\\
\hline
\label{tab:grid}
\end{tabular}
\end{table}

It is important to determine the radiative efficiency, $\eta$, 
in the SED fitting method. There are many approaches to estimate it. 
The $\eta \simeq 0.057$ computed for non-rotating BH.
\citet{Shankar09} suggest $\eta = 0.05-0.1$ in relation to AGNs. Observational constrains on growth of BHs made by {\citet{Yu}} give us reasonable argument that $\eta$ should be $\gtrsim 0.1$.  Even more, \citet{Cao08} proposed $\eta = 0.18$ for AGNs with BH masses above $10^9$~\Msun.  Performed analysis allows us to conclude that $\eta$ should be in the range of $0.15-0.20$. Those values are required to obtain the conformity of SMBH masses in LBQS if we use our SED fitting and single-epoch virial methods. Thus, we adopt the value of $\eta=0.18$ in relation to both types of quasars -- LBQS and WLQs.

We use a simple $\chi^2$ procedure to find the best-fit model and evaluate the quality of the fit.
It is based on directly matching the photometric points to the AD model. 
In our approach, we calculate $ \chi^2 = \sum_{i=1}^n (O_i - E_i)^2/\sigma_i$ 
for each quasar, where $O_i$ and $E_i$ are observed and modeled monochromatic luminosity
$L_{\lambda}$ which correspond to the $i$th photometric point, 
$\sigma_i$ is the observed error, and $n$ is the total number of observed data for the quasar.  
Satisfactory fits are defined as those showing reduced $\chi^2 \lesssim 5.5$.

In order to use information on yet determined black hole masses,
accretion rates, and their errors in our WLQs sample we carry out a
more sophisticated statistical analysis using the Bayesian method,
which is the conditional distribution of the uncertain quantity given
the data.  The values of BH masses and accretion rates (\mbhlit\ and
\dotmlit, respectively) of 9 WLQs were collected by different authors
(see Tab. \ref{tab:bestfit}, Col.~9) and those values for PG 1407 were
determined by us (see Subsec. \ref{sect:PG1407}). Note, that both
\mbhlit\ and \dotmlit\ of WLQs are based on the \FWHM(line)
determination. Authors use equations with factors suitable for normal
quasars which show strong lines and broad \FWHM. However, this is not
true for many WLQs.  For this reason, both values \mbhlit\ and
\dotmlit\ could be calculated wrongly.

The Bayesian inference method requires the knowledge of the prior
probability distribution $P(H|I)$, which represents our beliefs about
a hypothesis $H$ before some evidence, $I$, is taken into account. In
our calculations $H$ is $j$th model, $mod_j = mod(M_{\mathrm{BH} j},
\mdotm_j, a_{\ast j}, i_j)$, from among 366000 models we put in (note
that both of the analyses i.e. $\chi^2$ and the
Bayesian, are based on the same set of constructed grid of the AD
models).  Any prior information about this $j$th model is $I$. In
our case, the information $I$ should be \mbhobs, \dotmobs, etc., which are
observed. Nevertheless, we do not have those real parameters and
therefore $I$ is based on earlier calculated \mbhlit, \dotmlit\ and
their errors. Assuming a Gaussian probability distribution for
\mbhlit\ with standard deviations equal to $\sigma_M$, the prior can
be written as $P(H|\mmbhlit) \propto \exp{(-(\mmbh - \mmbhlit)^2/2\sigma_M^2)}$. The prior probability related to \dotm\ takes a similar Gaussian form. We do not have a prior knowledge on either BH spin and inclination. We assume delta function probability distribution for both parameters.

For each $j$th model, we also derive its likelihood function
$\mathcal{L}(mod_j) \propto \exp{(-\chi^2/2)} = P(D|mod_j,I)$, where
$D$ is the set of photometric points measured for each WLQ
quasar. Note, that there is no free parameters.

Finally, the posterior probability is determined for each model, as
the product of the likelihood and the priors on \mbh\ and
\dotm\ \citep[for details see][Appendix A]{Capellupo15}. It is given
by:
\begin{multline}
  P\left( H|D,I \right)=N \times \exp{\left( \frac{-\chi^2}{2}\right)} \times \\
  \times \exp{\left( -\frac{(\mmbh-\mmbhlit)^2}{2\sigma_M^2} \right)} \times \\
   \times \exp{\left( -{\frac{(\mdotm-\mdotmlit)^2}{2\sigma^2_{\dot{m}}}} \right)} 
\end{multline}
where $N$ is the normalization constant.

The Bayesian analysis identifies a model which has the highest probability
of explaining the observed SED assuming knowledge of BH mass and
accretion rate. We find that the number of sources with satisfactory
fit are the same and have high probability when we use the $\chi^2$
and Bayes' theorem.  The results for model with the highest posterior
probability are shown in Tab. \ref{tab:Bayes}.
Finally,  we take the masses \mbh\ calculated from the $\chi^2$ evaluation for further analysis.
\subsubsection{BH mass and accretion rate determination in PG~1407, errors estimation for Bayesian analysis}
\label{sect:PG1407}

There is a lack of the SMBH mass determination in the PG~1407 quasar.
We estimate it based on the equation (7.27) from \citet{Netzer13}, which states
$\mmbh \propto \left( \nu L_\nu (5100{\angstrom})
\right)^{0.65} \times \mFWHM(H \beta)^2$. 
The level of continuum at 5100{\angstrom} in PG~1407 is $\nu L_{\nu} = 3.16 \times 10^{46}$ \ergs\ \citep{Mcdowell95}. 
Unfortunately, the \Hb\ emission-line is almost undetectably weak \citep{Mcdowell95}. Firstly, 
we estimate \FWHM\ of \MgII\ line as follows. Using the \FeII\ template taken from
\citet{Vestergaard01} we subtract contribution of the
\FeII\ pseudo-continuum from magnesium line in the spectrum and we fit a Gauss function to it.  
The \FWHM(\MgII) calculated in this way is equal to $4300^{+1400}_{-530}$
\kms. Next, we convert the width of magnesium to
appropriate hydrogen based on the equation (6) by \citet{Wang09}:
$\log$(\FWHM(\MgII)) $\propto 0.81 \times \log(\mFWHM$(\Hb)).  Thus,
\FWHM(\Hb)$= 5400^{+2240}_{-810}$ \kms.
Finally, the calculated BH mass is $M_{\mathrm{BH}}^{\mathrm{H_{\beta}}} 
= (2.62^{+2.61}_{-0.73}) \times 10^9$~\Msun\ and we take this value as \mbhlit\ (Tab.~\ref{tab:bestfit}).
Additionally, we calculate BH mass based on \MgII\ line following the similar procedure. We use equation (7.28) from 
\citet{Netzer13} and get $M_{\mathrm{BH}}^{\mathrm{Mg(II)}} = (3.69^{+2.79}_{-0.85})\times\ 10^9$ \Msun.

The accretion rate in PG 1407 is calculated using relationship
$L_{Bol}/L_{Edd} \propto f(L) \times (L_{5100{\angstrom}})^{0.5}
\times (\mFWHM($\Hb$))^{-2}$ \citep[see equation 2 in][]{Plotkin15},
where $f(L)$ is the luminosity-dependent bolometric correction and equals to 5.7 
\citep{Shemmer10}. 
Eventually, $\mdotmlit = 0.45$ based on calculated \FWHM(\Hb) and
luminosity at 5100\AA\ in PG~1407.
\label{sect:Accrateerrors}

For Bayesian analysis we need to determine the errors of the accretion rates of those WLQs 
for which, the literature does not provide them. We use mentioned equation (2) \citep[see][]{Plotkin15}, 
upper and lower limits of \FWHM(\Hb) and $L_{5100{\angstrom}}$.
Errors of \dotmlit\ are listed in Tab. \ref{tab:bestfit}.

\section{Results}
\label{sec:4}

\begin{figure}
    \centering
    \includegraphics[width=0.480\textwidth]{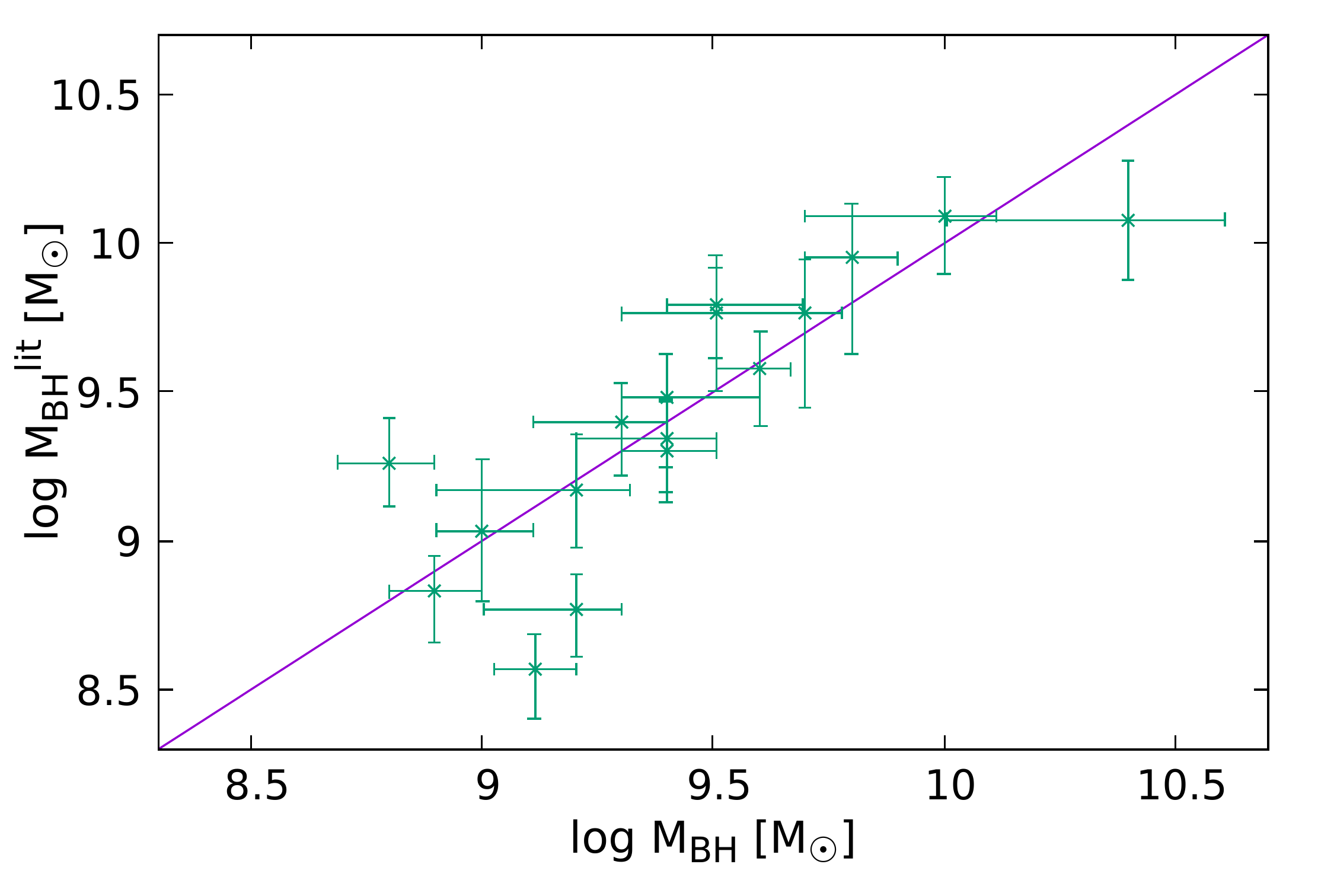}
    \caption{Comparison of LBQS masses (on the y-axis) with \mbh\ from our
      model (on the x-axis). Violet solid line is identity 1:1 line.}
    \label{fig:lbqs}
\end{figure}

For initial analysis, we use 27 quasars from the LBQS survey (see
Sec. \ref{sec:2}). Fig. \ref{fig:lbqs} shows us a comparison of
the supermassive black hole masses determined by
\citet{Vestergaard09}, \mbhlit\ (on the y-axis), to those obtained by
us, \mbh\ (on the x-axis). Both masses are given in mass units of the
Sun. Violet solid line is a 1:1 identity line.  \citet{Vestergaard09}
used the black hole mass determination based on their formula (1),
which is proportional to \FWHM (line) and luminosity $\nu L_{\nu}$. We
would like to note that we use the same grid of 366000 models (see
Sec. \ref{sec:3}) to obtain \mbh.  Compliance of masses and relatively
small distribution of errors means that the continuum fitting method
applies to quasars.

\begin{figure}
    \centering
    \includegraphics[width=0.48\textwidth]{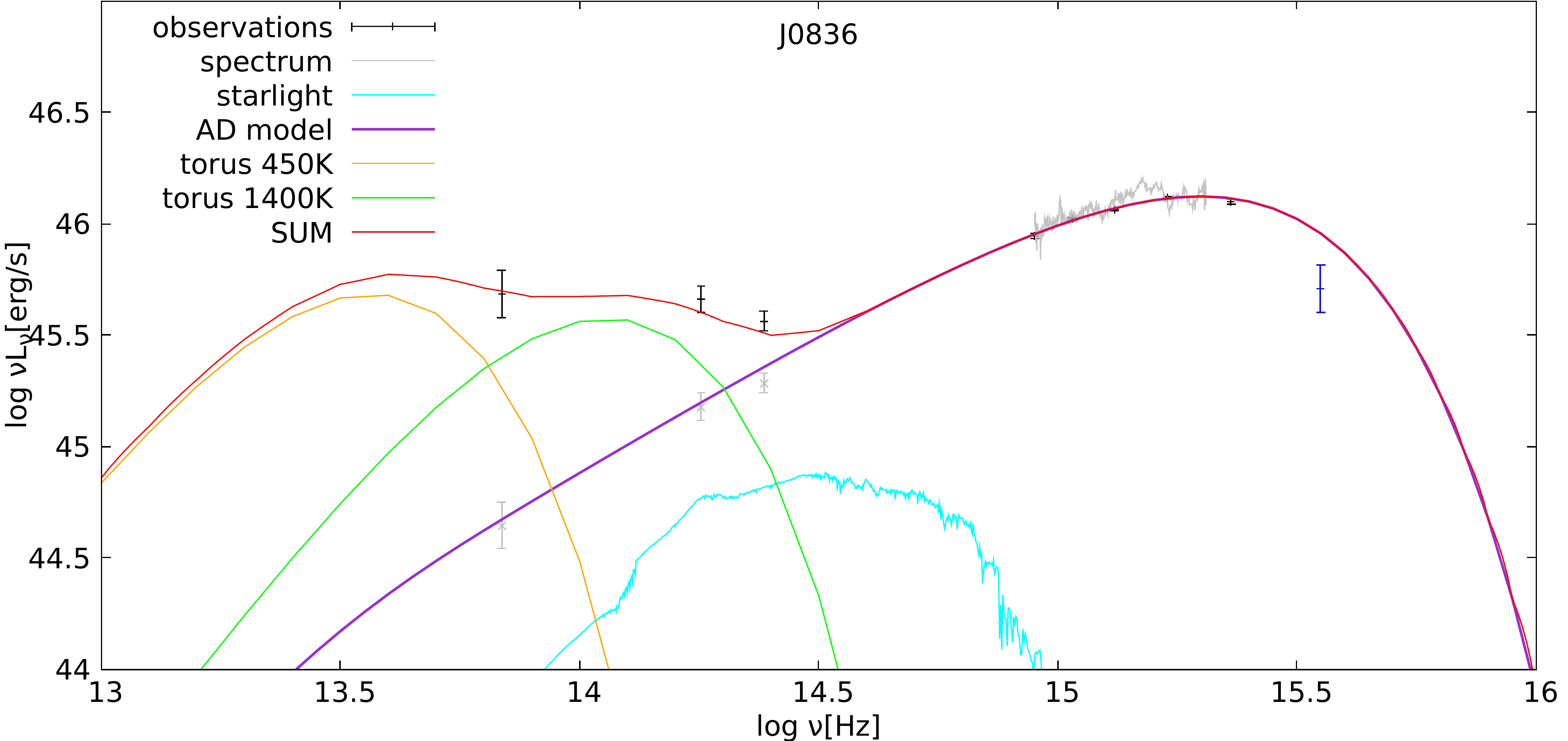}
    \caption{The best fit of WLQ SDSS J083650.86+142539.0. Summary of all components. Black points with errors show photometry data, grey line represents spectrum. Violet line shows contribution of the AD model. Tori with temperature 450K and 1400K, respectively, are displayed by orange and green lines. Cyan line points the level of starlight out. Sum of the components thus the best fit is shown by red line.}
    \label{fig:SUM}
\end{figure}

\begin{figure*}
     \begin{center}
            \includegraphics[width=0.35\textwidth]{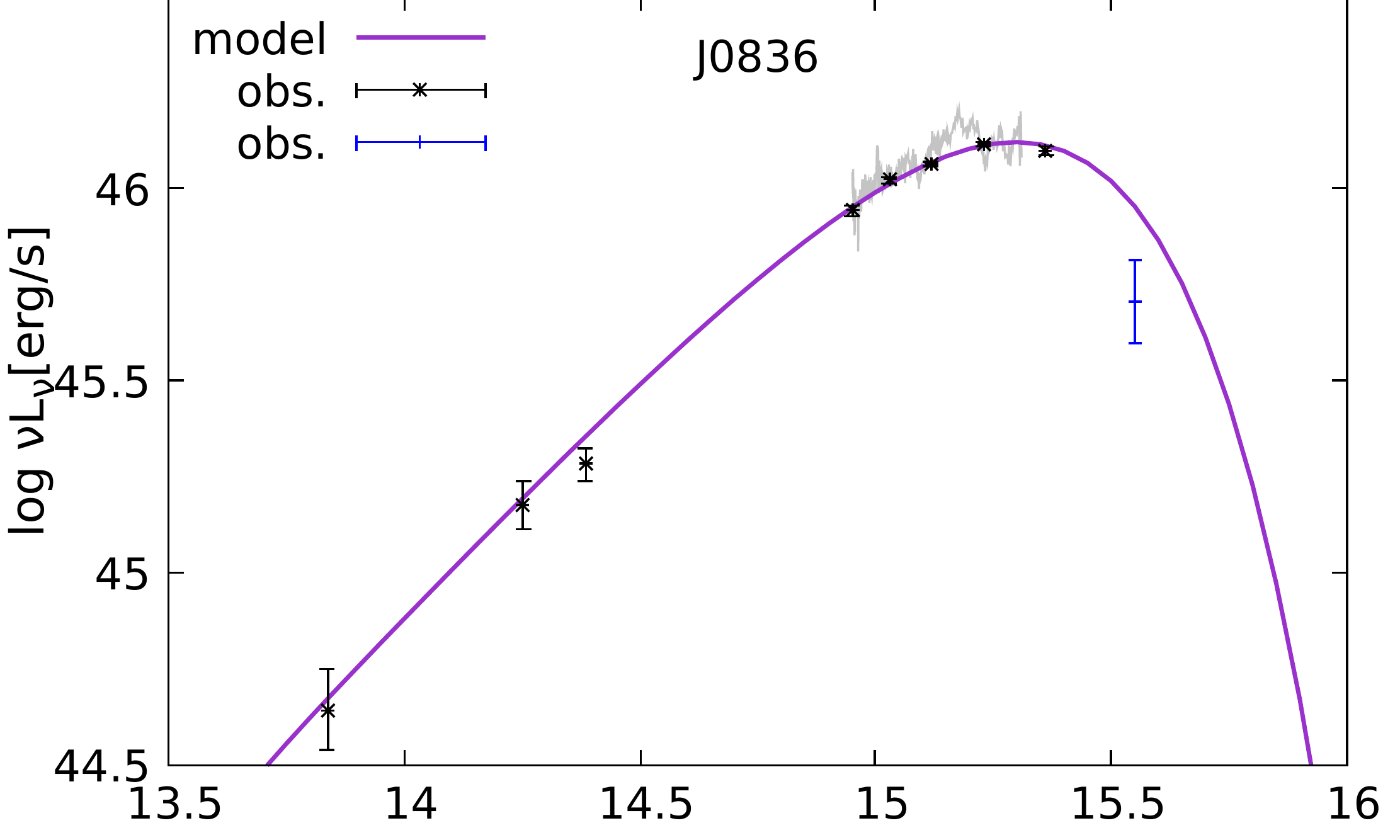}
            \includegraphics[width=0.35\textwidth]{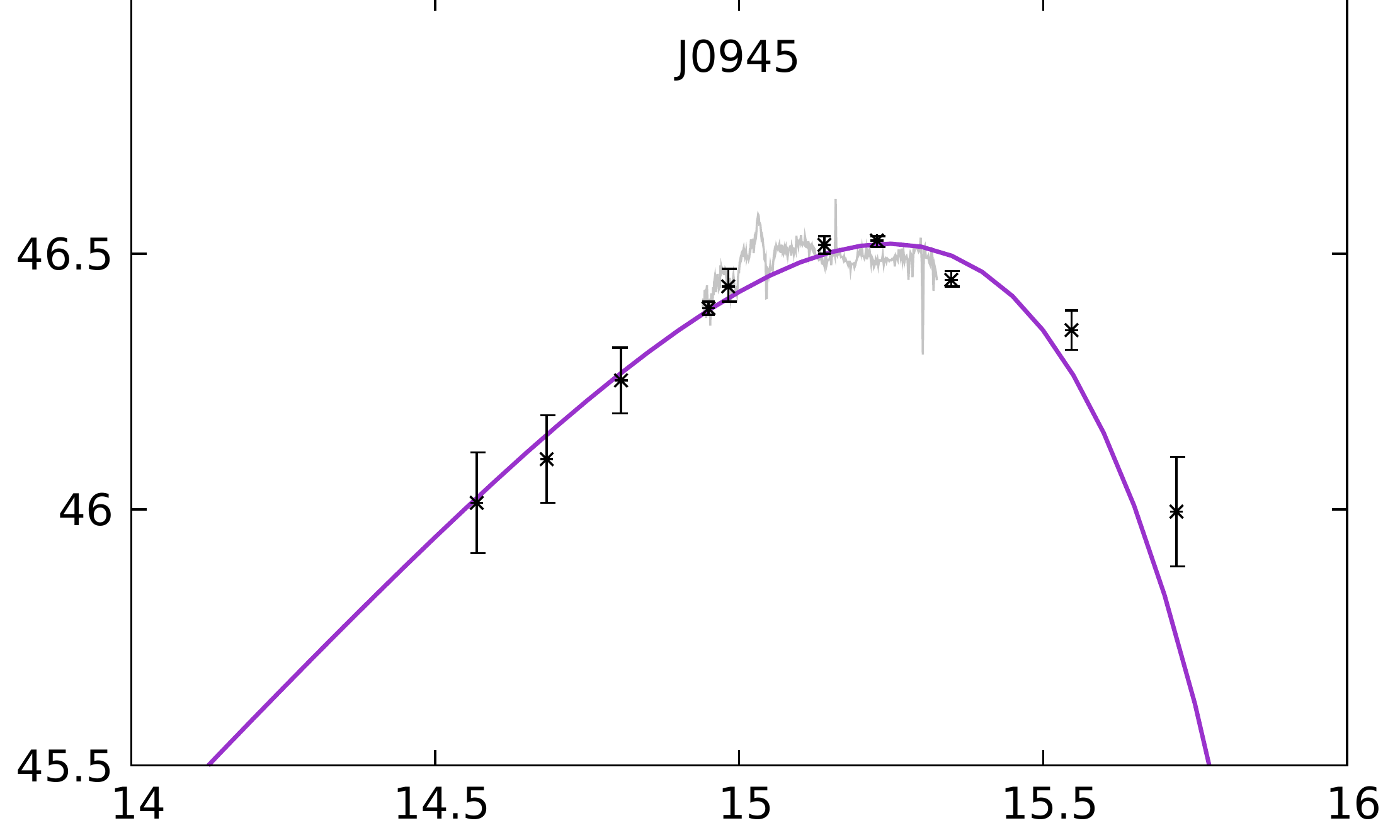}
            \includegraphics[width=0.35\textwidth]{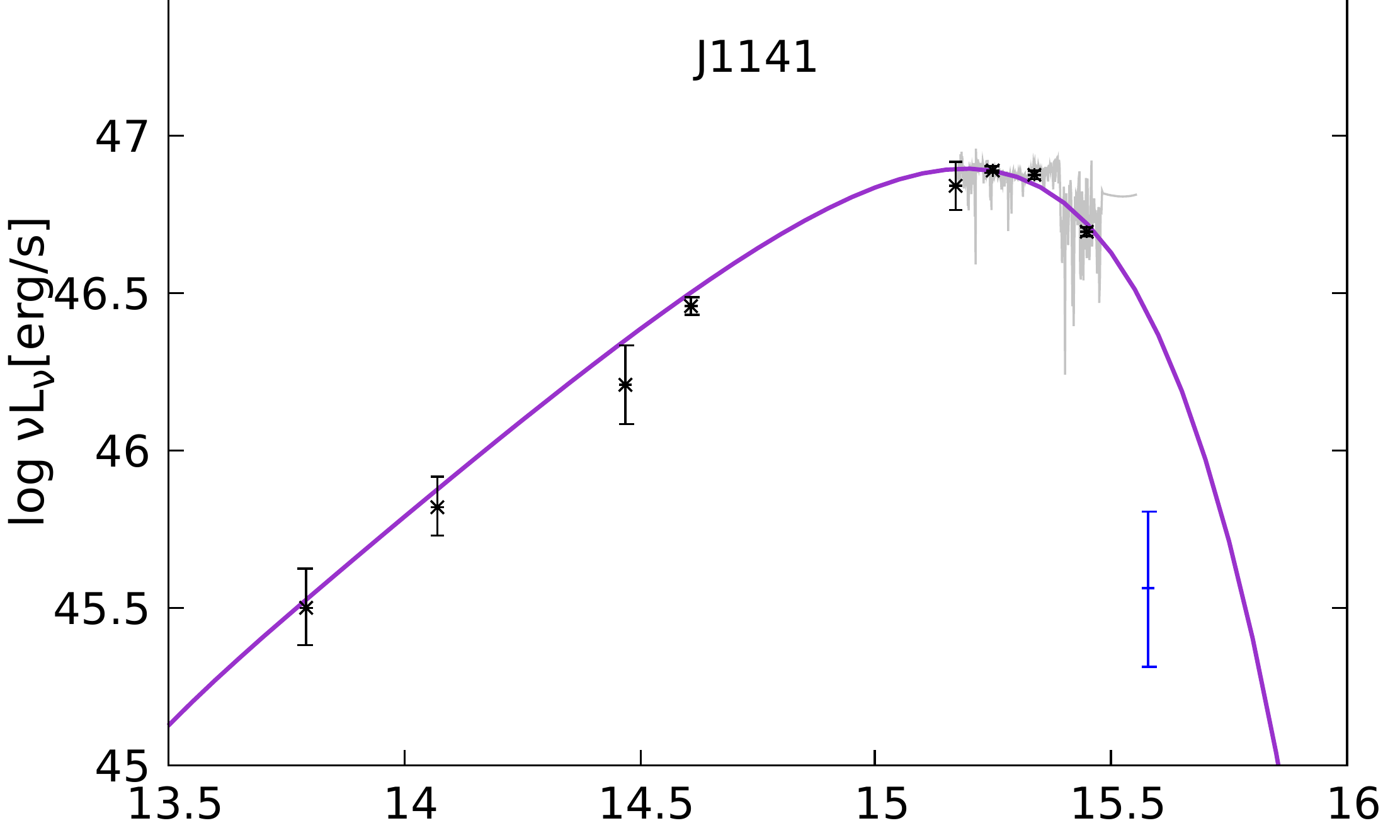}
            \includegraphics[width=0.35\textwidth]{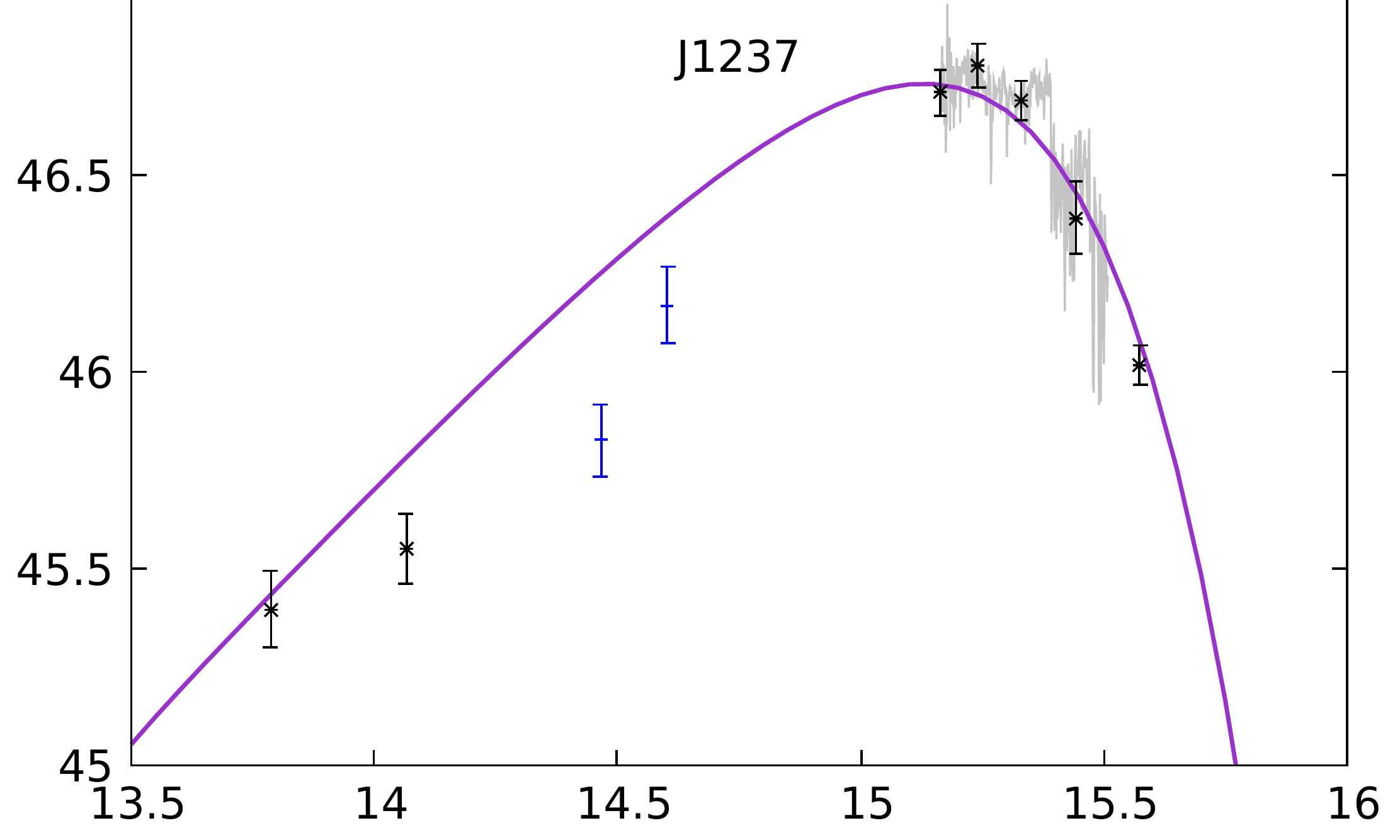}
            \includegraphics[width=0.35\textwidth]{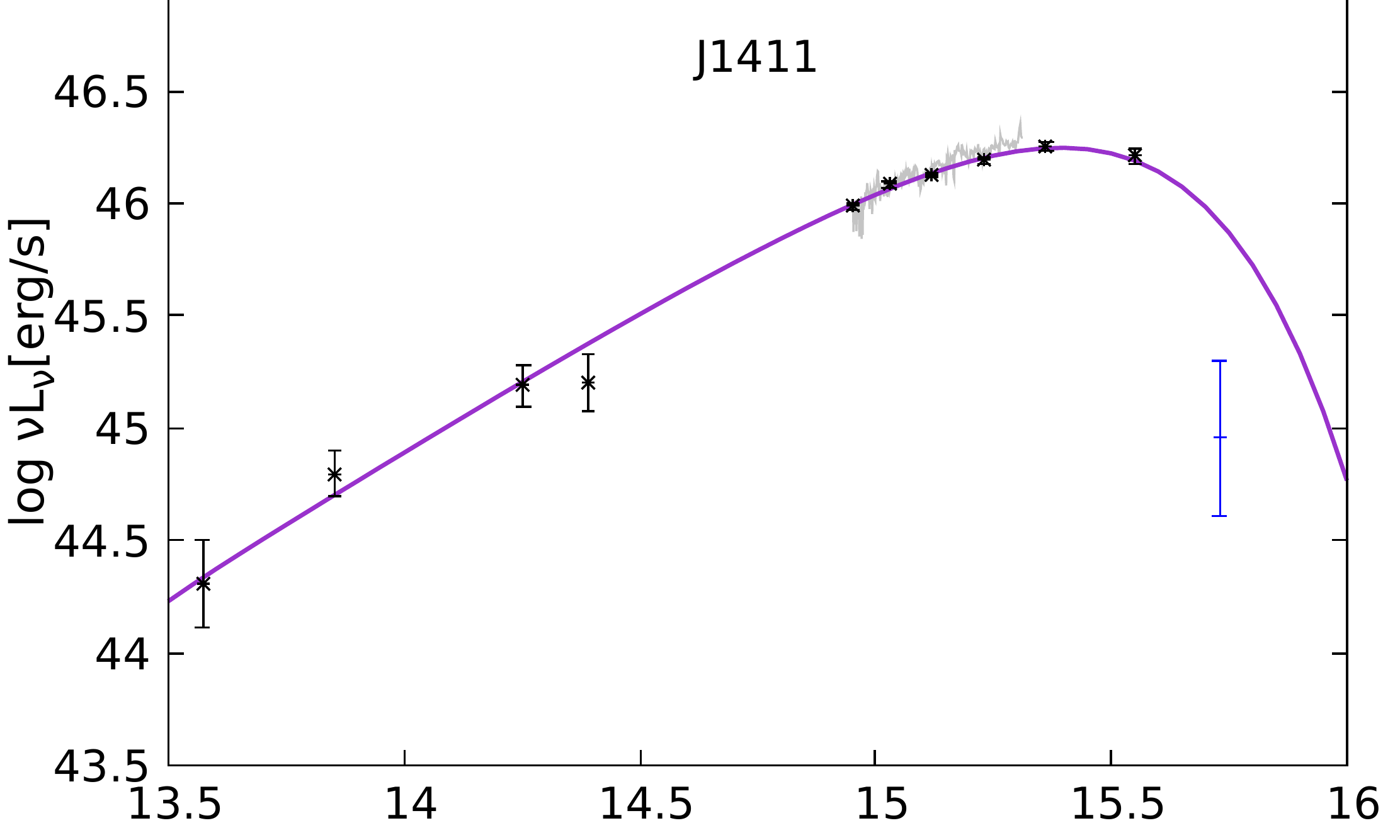}
            \includegraphics[width=0.35\textwidth]{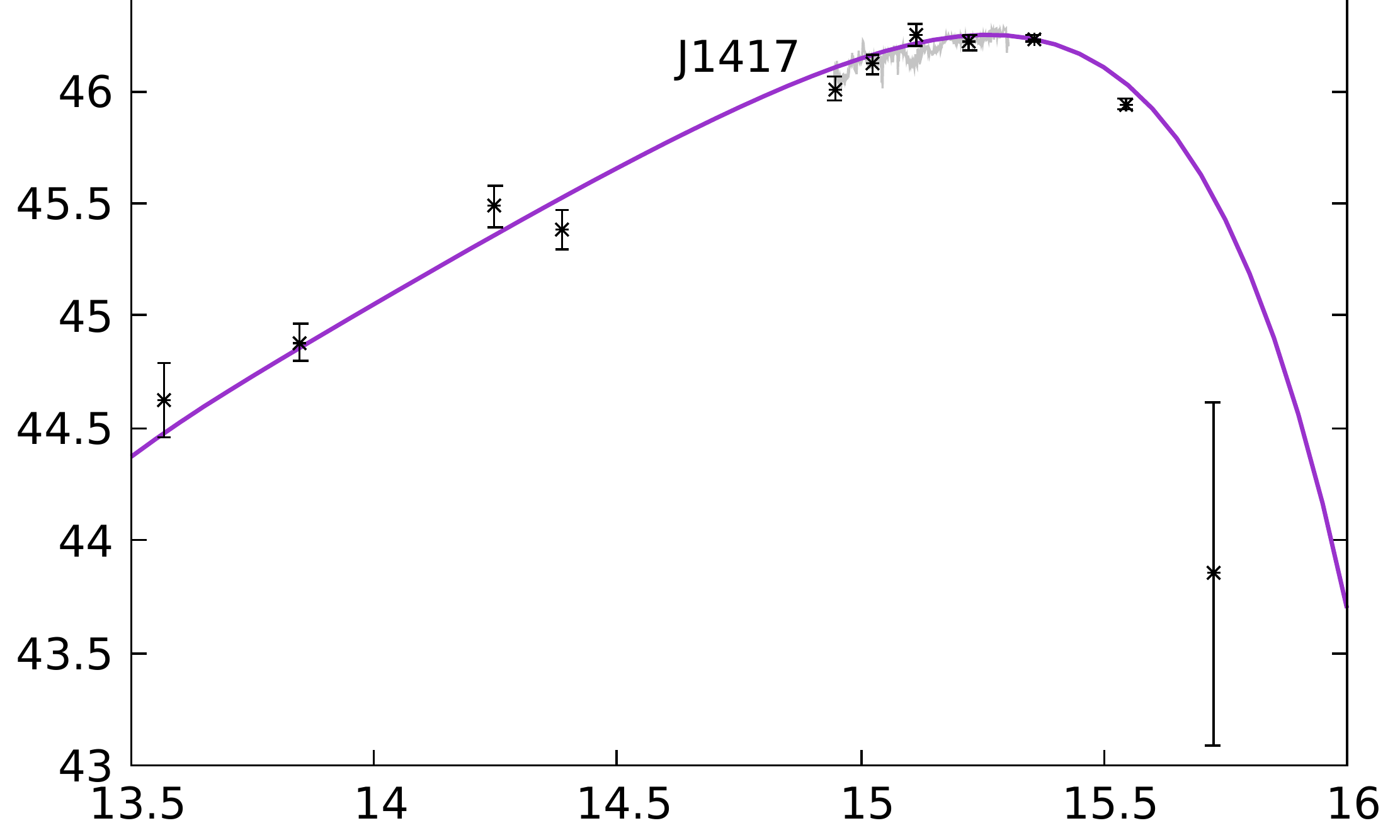}
            \includegraphics[width=0.35\textwidth]{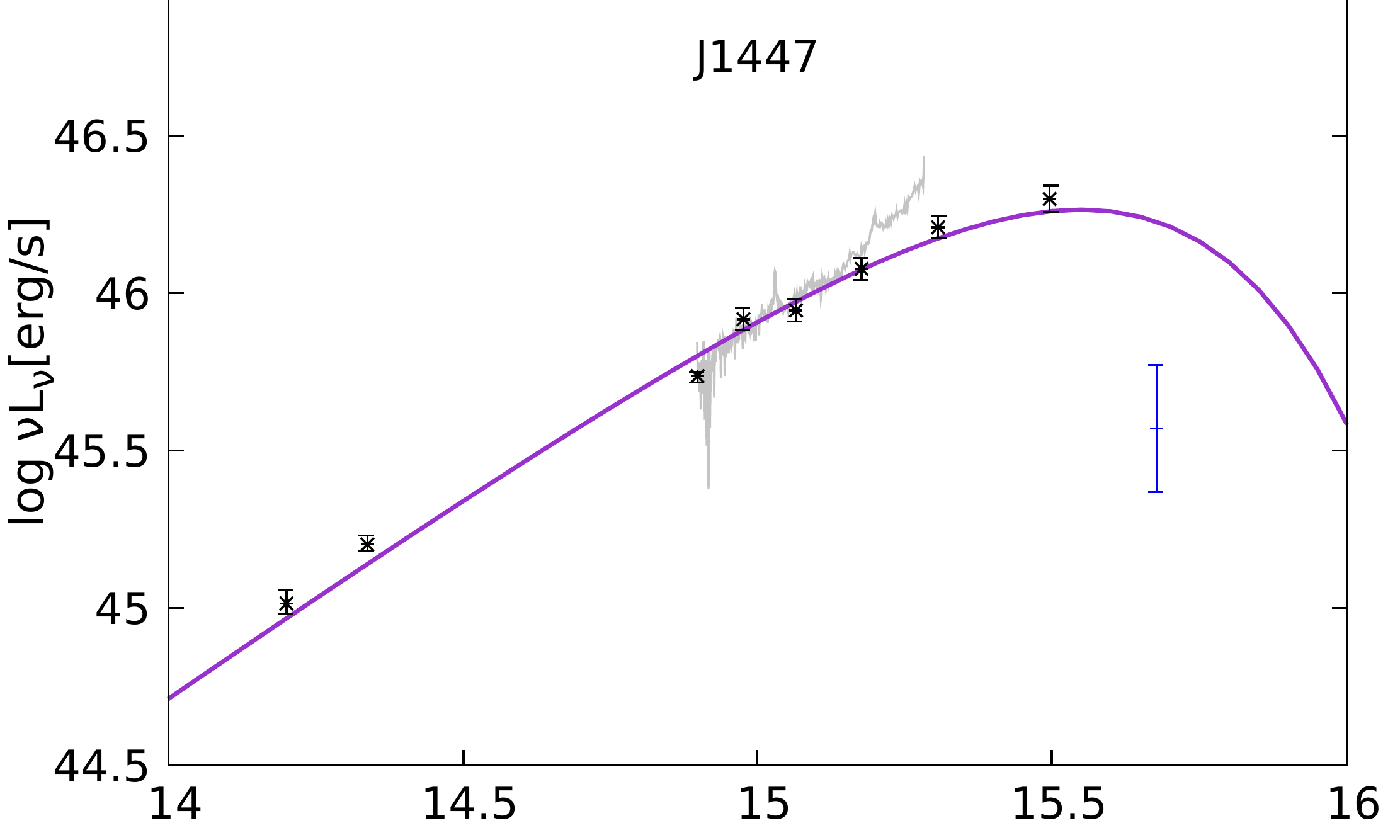}
            \includegraphics[width=0.35\textwidth]{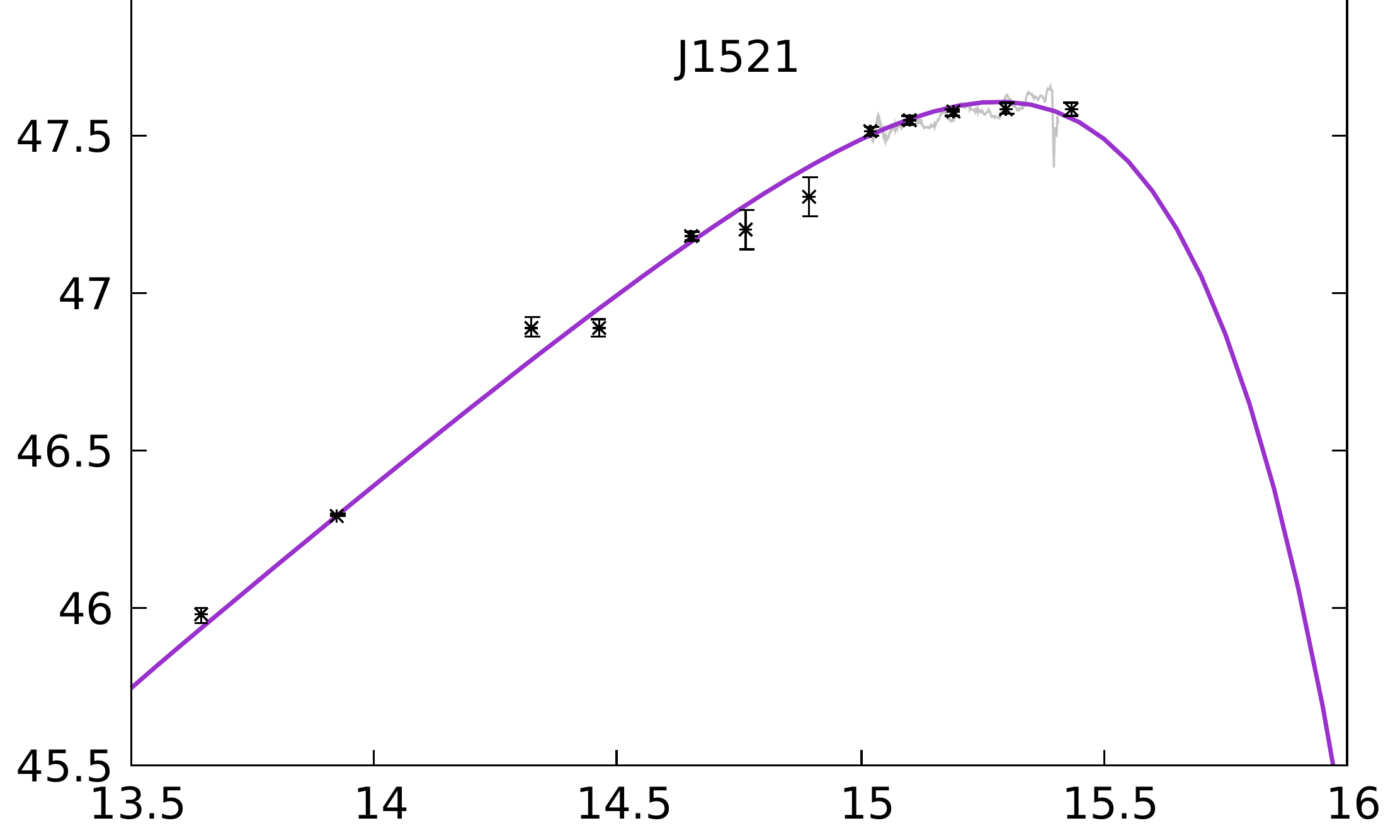}
            \includegraphics[width=0.35\textwidth]{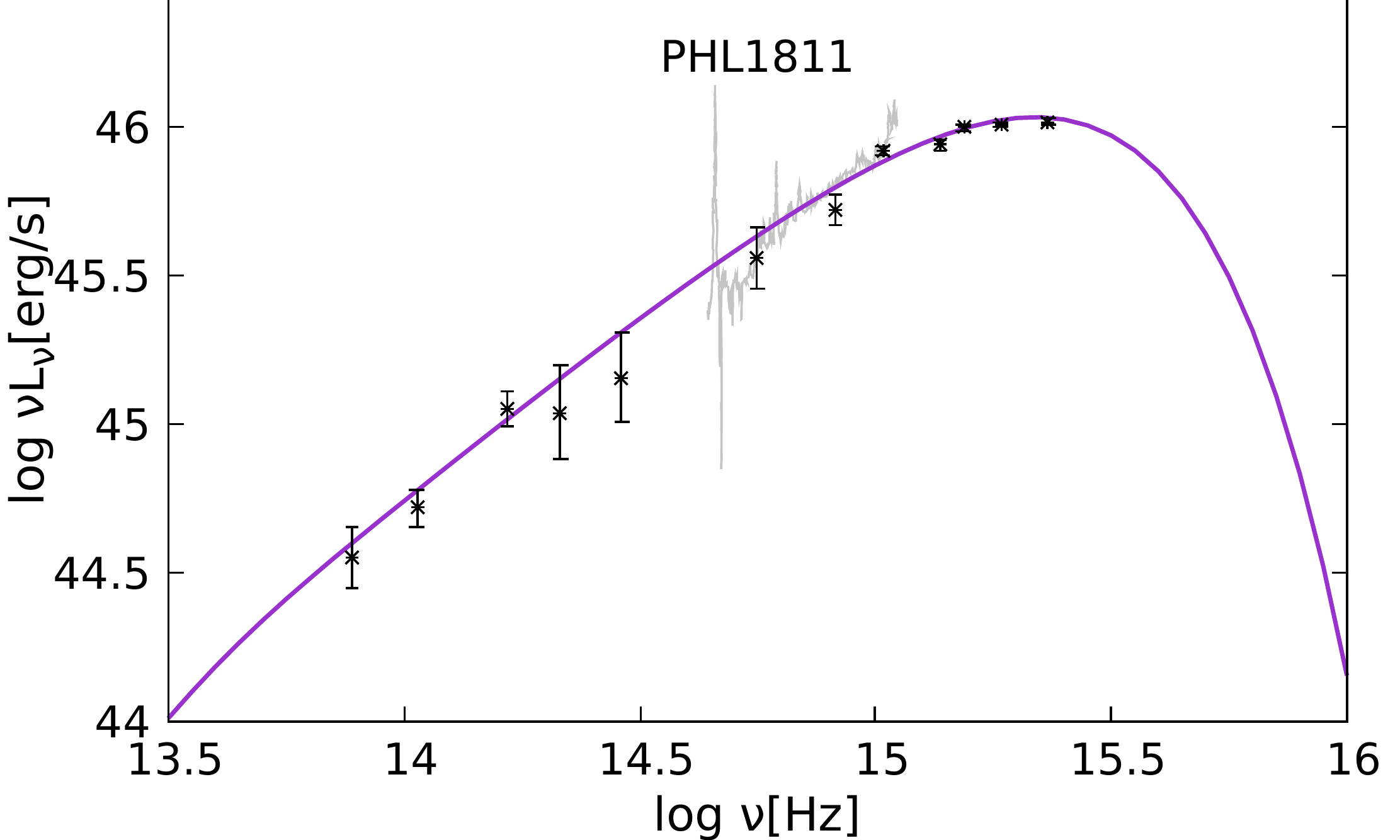}
            \includegraphics[width=0.35\textwidth]{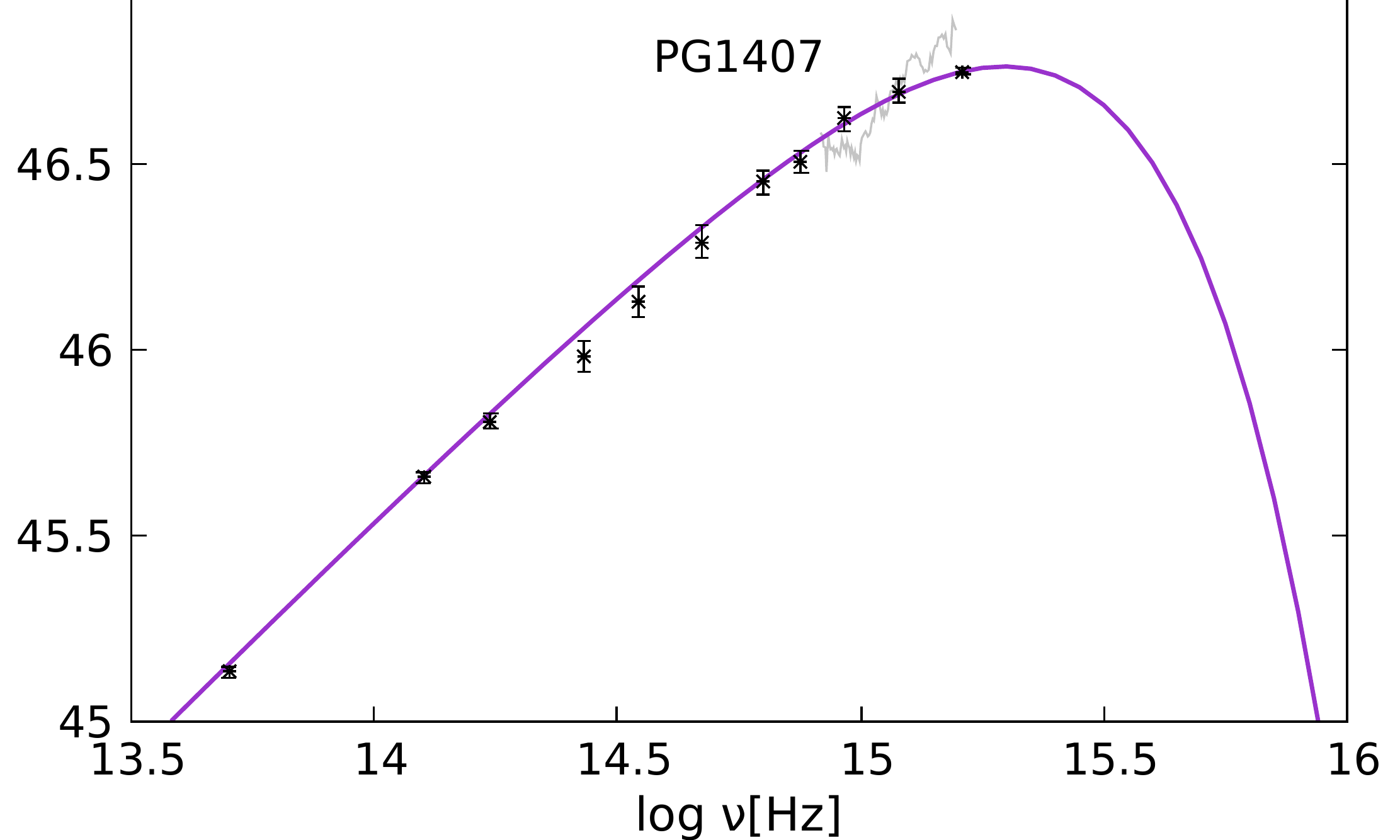}
    \end{center}
    \caption{The best fit of SED to photometric points of 10 WLQs. Black
      crosses and blue with errors show corrected observational data, grey line 
      -- spectra. Violet
      solid line represents the theoretical curve of the AD continuum model.}
   \label{fig:fit}
\end{figure*}

Our sample of weak emission-line quasars contains 10 objects.
Fig.~\ref{fig:SUM} shows in detail how the fitting procedure works. 
Different lines show the individual components: accretion disk, tori, 
starlight level. Solid red line shows sum of those components. 
The same approach is used in the rest of 9 WLQs.
Fig. \ref{fig:fit} presents the best fits of disk continua that
match the quasars SED. On the x-axis is the logarithmic value of
frequency in Hertz, while on the y-axis is the logarithmic value of
$\nu L_{\nu}$ in \ergs.  
The accretion disk continuum is marked with a solid purple line, the
photometric data are shown by black crosses and blue points with
errors.  The blue points observed in UV bandpass of 5 WLQs (J0836,
J1141, J1411, J1417, and J1447) and in optical range (J0836) could
suggest an absorption seen in some quasars (e.g. KVRQ 1500-0031
\citealt{Heintz18}, SDSS J080248.18+551328.9 \citealt{Ji15,Liu15}). The
absorption are caused by intrinsic gas in the host galaxy and/or
bigger influence of the assumed UV photoelectric absorption. Blue points
are outliers and for this reason we model the best fits in both cases: taking 
into consideration all points with and
without outliers ($\chi^2$ values in parenthesis of Tab.~\ref{tab:bestfit}).

\begin{deluxetable*}{lcccclccc}
\tablenum{4}
\tablecaption{The best fit parameters}
\tablehead{
\colhead{Name} & \colhead{\mbh} & \colhead{\dotm} & \colhead{\as} & \colhead{\incl} & \colhead{$\chi^2$/d.o.f} &
\colhead{\mbhlit} & \colhead{$\dot{m}_{\rm lit.}$} & \colhead{ref.} }
\decimalcolnumbers
\startdata
J0836 & $(1.30 ^{+5.00} _{-0.30}) \times 10^{9}$ & ${0.38 ^{+0.28}_{-0.25}}$ & ${0.00 ^{+0.30}_{-0.00}}$ & ${0.26 ^{+0.52}_{-0.26}}$ & 2.48 (1.91) & $(3.89^{+13.11}_{-1.75})\times 10^8$ & ${0.87 ^{+1.36}_{-0.65}}$ & 1/1 \\
J0945 & $(2.00 ^{+4.30} _{-0.50})\times 10^9$ & $0.47^{+0.19}_{-0.16}$ & ${0.00 ^{+0.30}_{-0.00}}$ & $0.52 \pm 0.26 $ & 3.73 & $(1.12 ^{+0.43}_{-0.19})\times 10^9$ & $0.51 \pm 0.15$ & 1/1 \\ 
J0945 &  &  &  &  & & $(3.08 \pm 0.27) \times 10^9 \dagger$ &  & 2 \\ 
J1141 & $(6.30 ^{+13.70} _{-1.30}) \times 10^9$ & $0.56^{+0.28}_{-0.36}$ & ${0.10 ^{+0.40}_{-0.10}}$ & $0.26 \pm 0.26$ & 4.58 (2.67) & $(3.16^{+1.41}_{-0.97}) \times 10^9$ & $0.38^{+0.20}_{-0.10}$ * & 3/3 \\  
J1237 & $(5.00 ^{+6.00} _{-1.80}) \times 10^9$ & $0.30^{+0.29}_{-0.21}$ & ${0.10}^{+0.60}_{-0.10}$ & ${0.26 ^{+0.52}_{-0.26}}$ & 6.64 (3.93) & $(2.00^{+1.39}_{-0.65})\times 10^9$ & $0.40^{+0.21}_{-0.16}$ * & 3/3 \\
J1411 & $(2.50 ^{+0.70} _{-1.20})\times 10^9$ & $0.31^{+0.44}_{-0.08}$ & ${0.80 ^{+0.10}_{-0.40}}$ & $0.26 ^{ +0.52}_{-0.26}$ & 5.58 (1.25) & $(5.25^{+3.87}_{-2.23})\times 10^8$ & $0.34^{+0.42}_{-0.17}$ & 1/1 \\
J1417  & $(3.20 ^{+0.70} _{-1.90})\times 10^9$ & $0.54^{+0.19}_{-0.41}$ & $0.20^{+0.30}_{-0.10}$ & $0.00 ^{ +0.52}_{-0.00}$ & 2.88 & $(3.55^{+3.53}_{-2.15})\times 10^8$ & $0.92 \pm 0.50$ & 1/1 \\
J1447 & $(1.30 ^{+1.20}_{-6.30}) \times 10^9$ & $0.35^{+0.41}_{-0.07}$ & $0.80 \pm 0.10 $ & $0.26 \pm 0.26$ & 5.37 (4.41) & $(1.12^{+2.51}_{-1.25})\times 10^8$ & $1.30 ^{+0.17}_{-0.78}$ & 1/1 \\
J1521 & $(2.00^{+0.50}_{-5.70})\times 10^{10}$ & $0.48^{+0.19}_{-0.35}$ & ${0.80 ^{+0.10}_{-0.20}}$ & $0.52^{+0.26}_{-0.52}$ & 3.68 & $(6.20^{+1.73}_{-1.51}) \times 10^9$ & $0.81^{+0.27}_{-0.17}$ * & 4/4 \\
PHL 1811 & $(7.90^{+2.10}_{-3.90})\times 10^8$ & $0.34^{+0.50}_{-0.03}$ & $0.00^{+0.10}_{-0.00}$ & $ 0.00^{+0.26}_{-0.00}$  & 1.87 & $(1.14^{+2.57}_{-2.57})\times 10^8$ & $1.30^{+0.03}_{-0.02}$ * & 5/5 \\
PG 1407 & $(7.90 ^{+5.10}_{-2.90})\times 10^9$ & $0.26^{+0.28}_{-0.09}$ & $0.90^{+0.00}_{-0.10}$ & $0.78 \pm 0.26 $ & 1.42 & $(2.62^{+2.61}_{-0.73})\times 10^9$ & $0.45^{+0.17}_{-0.23}$ * & a/a \\
PG 1407 &  &  &  &  & & $(3.69^{+2.79}_{-0.85})\times 10^9 \dagger$ &  & a \\ 
\enddata
\label{tab:bestfit}
\tablecomments{ 
Black hole masses, Eddington accretion rates, spins, and fitted inclinations are in Col. (2)-(5), respectively. Col. (6) contains the normalized $\chi^2$ values in two cases: numbers without parenthesis -- all photometric points are taken into account, numbers in parenthesis -- data with removed outliers (only black points in Fig.~\ref{fig:fit}) are considered. The values in Col. (2)-(5) refer to the case where all points are fitted. The values of the parameters in the absence of outliers are the same as before within the errors. Black hole masses and accretion rates taken form literature are in Col.(7)-(8). They are based on \FWHM(\Hb) measurements (values without $\dagger$). $\dagger$ -- \mbh\ is based on \MgII\ line. \mbh\ and \mbh$_{\rm ,lit.}$ are in units of $M_{\odot}$. * -- errors of $\dot{m}_{\rm lit.}$ are estimated by us. Numbers refer to articles: 1) \cite{Plotkin15}, 2) \cite{Hryniewicz10}, 3) \cite{Shemmer10}, 4) \cite{Wu11}, 5)
  \cite{Leighly07b}, 6) \cite{Mcdowell95}, a) this work.}
\end{deluxetable*}

\begin{table}
    \tablenum{5}
    \centering
    \caption {The Schwarzschild black hole solutions (\as=0)}
    \begin{tabular}{ccccc}
    \hline   
    Name & \mbh& \dotm & \incl & $\chi^2$/d.o.f\\
    \hline
    J0836 & $(1.30 ^{+5.00} _{-0.30}) \times 10^{9}$ & ${0.38 ^{+0.28}_{-0.25}}$ & ${0.26 ^{+0.52}_{-0.26}}$ & (1.91)\\
    J0945 & $(2.00 ^{+4.30} _{-0.50})\times 10^9$ & $0.47^{+0.19}_{-0.16}$  & $0.52 \pm0.26$ & (3.73)\\
    J1141 & $(7.90 ^{+2.10} _{-2.90}) \times 10^{9}$ & ${0.84 ^{+0.03}_{-0.28}}$ & ${1.30 ^{+0.00}_{-0.52}}$ & (4.11)\\
    J1237 & $(4.00 ^{+2.30} _{-0.80}) \times 10^{9}$ & ${0.30 ^{+0.06}_{-0.29}}$ & ${0.00 ^{+0.26}_{-0.00}}$ & (4.54)\\
    J1411 & $(7.90 ^{+3.40} _{-1.60}) \times 10^{9}$ & ${0.50 ^{+0.03}_{-0.11}}$ & ${0.00 ^{+0.26}_{-0.00}}$ & (1.43)\\
    J1417 & $(2.50 ^{+0.70} _{-0.50}) \times 10^{9}$ & ${0.66 ^{+0.19}_{-0.22}}$ & ${1.30 ^{+0.00}_{-0.26}}$ & (1.98)\\
    J1447 & $(5.00 ^{+2.90} _{-1.00}) \times 10^{9}$ & ${0.49 ^{+0.20}_{-0.05}}$ & ${0.78 ^{+0.52}_{-0.26}}$ & (4.96)\\
    J1521 & $(1.30 ^{+0.70} _{-3.40}) \times 10^{10}$ & ${0.86 ^{+0.11}_{-0.63}}$ & ${1.30 ^{+0.00}_{-0.52}}$ & (3.99)\\
    PHL 1811 & $(7.90^{+2.10}_{-3.90})\times 10^8$ & $0.34^{+0.50}_{-0.03}$ &
    $ 0.00^{+0.26}_{-0.00}$  & (1.87)\\
    PG 1407 & $(2.50 ^{+0.70} _{-0.50}) \times 10^{9}$ & ${0.85 ^{+0.06}_{-0.19}}$ & $0.78\pm 0.26$  & (0.93)\\
 \hline
 \label{tab:Schwarzschild}
    \end{tabular}
 \tablecomments{
The normalized $\chi^2$ values in parenthesis means the data with removed outliers (only black points in Fig.~\ref{fig:fit}) are considered.}
 \end{table}
Our results are collected in Tab. \ref{tab:bestfit}. The black hole masses, 
accretion rates, spins, inclinations of each WLQs,
and $\chi^2$ values are in Columns (1)-(6). Columns (7) and (8)
contain the literature values of the black hole masses and the
accretion rates, respectively. Column (9) contains references to the
above values, as appropriate. 
Degeneration of solutions due to the spin parameter takes place.
Two groups of the best fit for zero and non-zero spin are difficult to distinct.
Thus, we  also perform additional fit with the fixed \as\ equals 0  (Tab.~\ref{tab:Schwarzschild}).
In this approach, the photometric points without outliers are taken into account (only black points in Fig.~\ref{fig:fit}).

Additionally, the black hole masses determined from the Bayesian analysis are presented in
Tab. \ref{tab:Bayes}. No significant differences in black hole masses calculated from both the $\chi^2$ and the Bayesian analysis
are indicated. It suggests that the determined global parameters are correct and describe
the overall SED shape of these objects. We take the values from Tab.~\ref{tab:bestfit} for further analysis.

\begin{figure}
    \centering
    \includegraphics[width=0.480\textwidth]{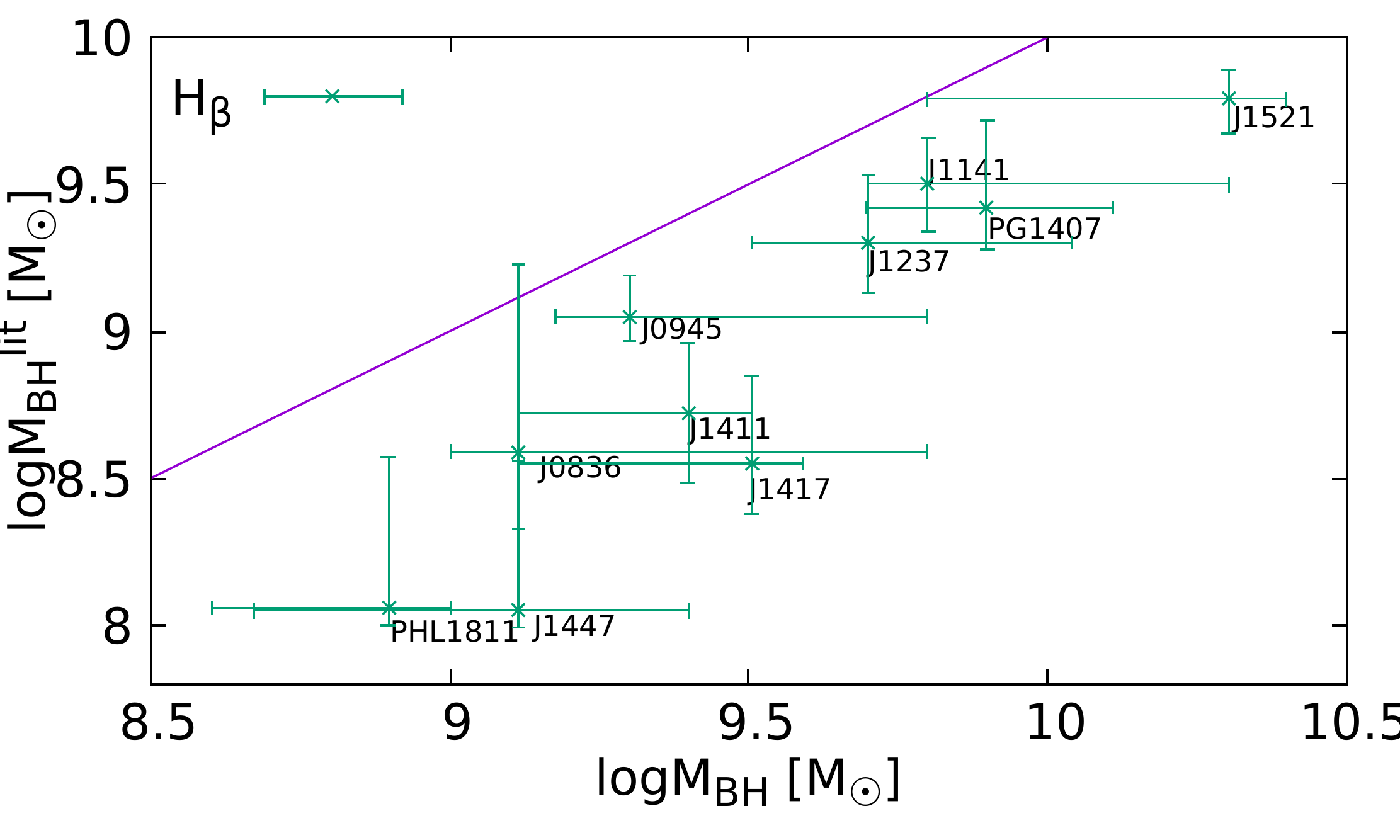}
    \caption{Comparison of SMBH masses (\mbhlit) of WLQs based on
      \FWHM\ estimations with \mbh, that come from spectral fitting
      method. Violet solid line is identity 1:1 line.  Green crosses
      represent \mbhlit\ calculated using \FWHM\ of \Hb.}

    \label{fig:Comparison}
\end{figure}

\begin{figure}
    \centering
    \includegraphics[width=0.480\textwidth]{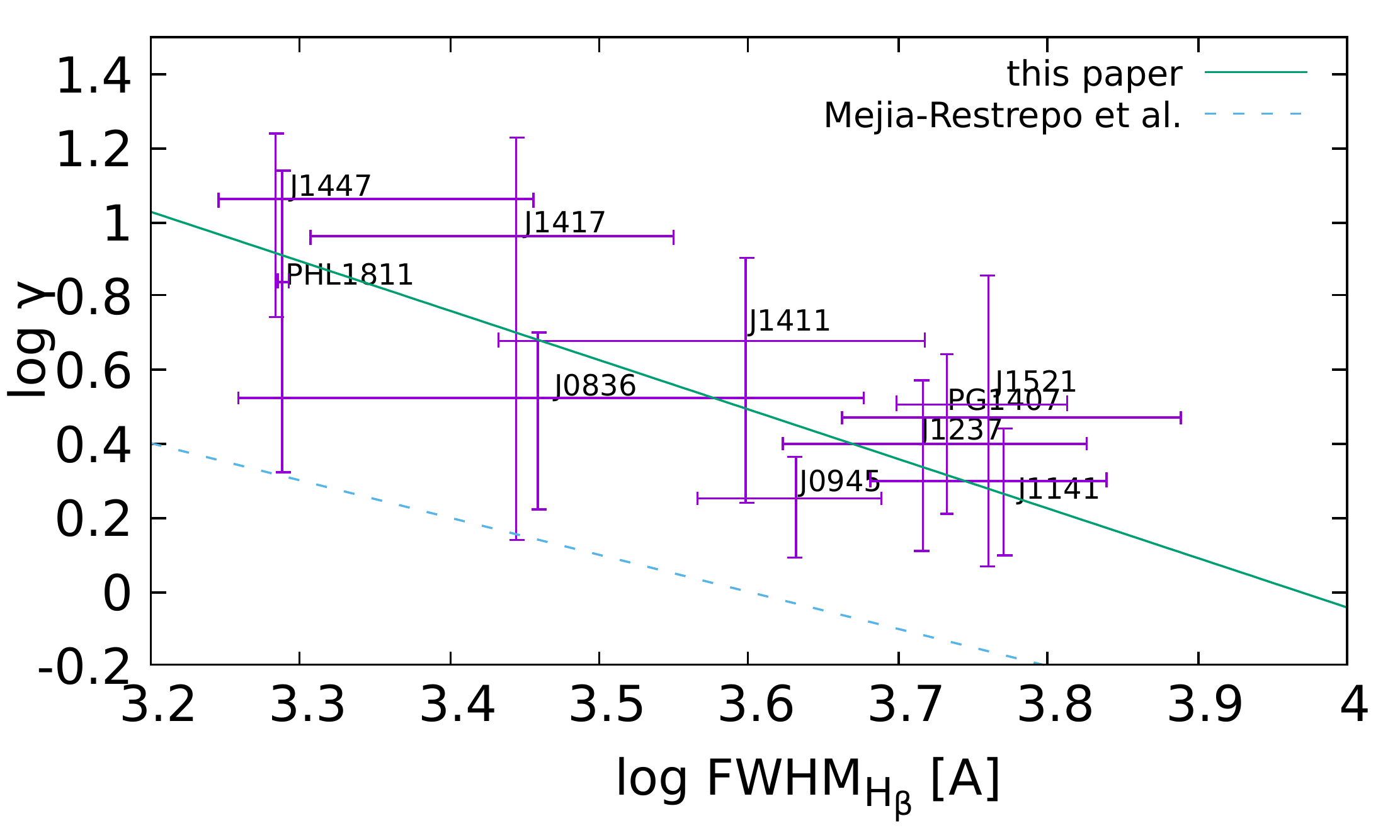}
    \caption{The $\gamma$ factor ($= \mmbh/\mmbhlit$) versus
      \FWHM(\Hb). The best fit (i.e. $\gamma \propto \mFWHM^{-1.34}$)
      is shown by green solid line. Points show data of 10 WLQs.  The
      blue dash line represents the fit of the virial factor (i.e. $f
      \propto \mFWHM^{-1.17}$) to 37 AGNs made by \citet{Mejia19}. In
      our case $\gamma = \mathrm{const} \times f$ and the dash line is
      shifted down.}
    \label{fig:factor+}
\end{figure}

Fig.~\ref{fig:Comparison} shows 
the mass distribution of black holes. Identity 1:1
line is marked as solid purple line. The presented mass comparison suggests
that literature determinations of black hole masses, \mbhlit, based on
\FWHM(\Hb) are generally underestimated.  We also determine the
difference between \mbhlit\ and \mbh\ values. The $\gamma$ factor is
calculated ($\gamma = \mmbh/\mmbhlit$). In Fig.~\ref{fig:factor+} we
present the relationship between the logarithmic value of \FWHM(\Hb)
in \kms\ (see Tab. \ref{tab:fwhm and gamma}) and the logarithmic value
of the $\gamma$ factor. The green solid line shows the best fit
between those variables. The fit is made using the nonlinear
least-squares (NLLS) Marquardt-Levenberg algorithm which takes into
account errors in both $x$ and $y$ directions.  The relationship is:
\begin{multline}
            \log \gamma =\left(-1.338 \pm 0.366\right) \times 
            \log \left( \frac{\mFWHM (H_{\beta})}{10^3 \, \mkms} \right)+ \\
        \left(1.294 \pm 0.234\right) 
        \label{eq:gamma_eq}
\end{multline}

For a better assessment of our calculations, we have determined
the Spearman coefficient, which is $r_{s} = -0.806$ and the linear correlation coefficient, $r = -0.82$. It can be seen that three of the ten objects (J0945, J1141, and J1237) are close to 1:1 line and the $\gamma$ factor is $\lesssim 2.5$. Masses of three other sources (PHL 1811, J1417, J1447)  should be multiplied by $\gamma > 7$. The mean $\gamma$ factor is 4.7, and median is 3.3. The dashed blue line in Fig. \ref{fig:factor+} represents the best fit obtained by \citet{Mejia19}. They used a sample of 37 Type I AGNs, which lie in the range of redshifts $\sim1.5$. Eq. (\ref{eq:gamma_eq}) allows us to correct the black hole masses, \mbhlit, determined so far and based on \FWHM\ values of \Hb\ line.

   \begin{table}
   \tablenum{6}
    \centering
    \caption {The BH masses from the Bayesian analysis}
    \begin{tabular}{cc}
    \hline   
    Name & \mbhbay\ [\Msun]\\
    \hline
    J0836 & $(2.00^{+4.30}_{-1.00})\times 10^9$ \\
    J0945 & $(3.20^{+3.10}_{-1.20})\times 10^9$ \\
    J1141 & $(7.90^{+4.10}_{-2.90})\times 10^9$ \\
    J1237 & $(6.30^{+3.70}_{-1.30})\times 10^9$ \\
    J1411 & $(1.30^{+1.90}_{-1.30})\times 10^9$ \\
    J1417 & $(2.50^{+2.50}_{-1.20})\times 10^9$ \\
    J1447 & $(1.00^{+1.50}_{-6.00})\times 10^9$ \\
    J1521 & $(1.30^{+1.20}_{-5.00}) \times 10^{10}$ \\
    PHL 1811 & $(7.90^{+3.10}_{-3.90}) \times 10^8$ \\
    PG 1407 & $(7.90^{+3.40}_{-2.90}) \times 10^9$ \\
    \hline
    \label{tab:Bayes}
    \end{tabular}
   \end{table}
    \begin{table}
   \tablenum{7}
    \centering
    \caption {The Full Width at Half Maximum of \Hb\ emission-line and the $\gamma$ factor}
    \begin{tabular}{cccc}
    \hline   
    Name & \FWHM(\Hb) [\kms] & $\log \gamma$ & ref. \\
    \hline
    J0836 & $2880^{+1877}_{-1069}$ & $0.52^{+0.18}_{-0.31}$ & 1\\
    J0945 & $4278 \pm 598$ & $0.25^{+0.11}_{-0.16}$ & 1\\
    J1141 & $5900^{+1000}_{-1100}$ & $0.30^{+0.14}_{-0.20}$ & 2\\
    J1237 & $5200^{+1500}_{-1000}$ & $0.40^{+0.17}_{-0.29}$ & 2\\
    J1411 & $3966 \pm 1256$ & $0.68^{+0.22}_{-0.44}$ & 1\\
    J1417 & $2784 \pm 759$ &  $0.96^{+0.27}_{-0.82}$ & 1\\
    J1447 & $1923^{+933}_{-164}$ & $1.06^{+0.18}_{-0.32}$ & 1 \\
    J1521 & $5750 \pm 750$ * & $0.51^{+0.36}_{-0.44}$ & 3\\
    PHL 1811 & $1943\pm 19$ & $0.84^{+0.3}_{-0.52}$ & 4\\
    PG 1407 & $5400^{+2240}_{-810}$ $\sharp$ & $0.47^{+0.17}_{-0.26}$ & a\\
    \hline
    \label{tab:fwhm and gamma}
    \end{tabular}
    \tablecomments{ $\gamma$ = \mbh /\mbhlit. * -- errors are estimated by us. $\sharp$~\Hb\ is weak and almost not visible, its \FWHM\ is estimated based on \FWHM(\MgII). Numbers refer to \FWHM\ sources: 1) \cite{Plotkin15},  2) \cite{Shemmer10}, 3) \cite{Wu11}, 4) \cite{Leighly07b}, a) this work.}
    \end{table}

\section{Discussion}
\label{sec:5}

The virial factor ($f$ in Eq.~\ref{eq:blr}) is often assumed to be
constant with values of 0.6-1.8 \citep[e.g.][]{Peterson04RM, Onken04, Nikolajuk06}, where 0.75 corresponds to a spherical geometry of the BLR.  Generally, $f$ dependents on non-virial velocity components such as winds, the relative thickness $(H/\mRblr)$ of the Keplerian BLR orbital plane, the line-of-sight inclination angle ($i$) of this plane, the radiation pressure \citep{Wills86, Gaskell09, Denney09, Denney10, Shen14, Runnoe14} and it should be a function of those phenomenons. The analysis carried out by \citet{Mejia19} indicates a low influence of radiation pressure on the $f$ factor, however this mechanism cannot be excluded.  Whether or not we skip the radiation pressure influence, the line-of sight inclination of gas in a planar distribution of the BLR plays important role in black hole mass calculations. Unfortunately, the nature of the velocity component responsible for the thickness of the BLR and thus its geometry is unclear (e.g. \citealt{Done96,Collin06,Czerny16}; for recent review see \citealt{Czerny19}).  

Our results support those got by \citet{Mejia19}. They study 37 AGNs
at redshifts $\sim 1.5$.  The authors indicate the dependency of the
virial factor, on observed \FWHM\ of the broad emission-line (such as
$H\beta$, \MgII, \CIV) in the form of an anti-correlation.  It implies
that the BH mass estimations based on the reverberation or the single-epoch virial BH mass
method are systematically
overestimated for AGN systems with larger \FWHM\ (e.g. $\gtrsim 4000$
\kms\ for $H\alpha$) and underestimated for systems with small
$\mFWHM(H\alpha)\lesssim 4000$\kms.  It is worth to note that the
opposite rule applies to the Eddington accretion rates (because
$\mdotm \propto \mmbh^{-1}$).  
We found a similar underestimation of \mbhlit\ values in the
sample of WLQs (Fig.~\ref{fig:factor+}). 
SMBH masses of AGNs, which show \FWHM(\Hb)$\gtrsim 5000\ \mkms$ need to by multiplied 
by a small factor of 1.5-2.5 while the rest of them requires the larger factor up to 12.
It means that the masses of about 50-60\% of WLQs 
are underweight based on \FWHM(\Hb) values.

It can be seen in Fig.~\ref{fig:Comparison}
that the correction of the SMBH masses based on \FWHM\ estimation is
needed.  We modify \mbhlit\ (Equation 1 in \citealt{Plotkin15}):
\begin{equation}
    \frac{\mmbhlit}{10^6\mMsun} = 5.05 \left(\frac{\nu L_{\nu}(5100)}{10^{44} \mergs}\right)^{0.5}\left(\frac{\mFWHM(H_{\beta})}{10^3 \mkms}\right)^2
\end{equation}
using the definition of the $\gamma$ factor ($\mmbh=\gamma \times \mmbhlit$)
and Eq.~(\ref{eq:gamma_eq}).
The corrected formula for the SMBH masses in WLQs is:
\begin{multline}
    \frac{\mmbh}{10^7  \mMsun}=  (9.94^{+7.09}_{-4.13}) \times
    \left(\frac{\nu L_{\nu}(5100)}{10^{44} \mergs}\right)^{0.5} \times \\
    \times \left(\frac{\mFWHM(H_{\beta})}{10^3 \mkms}\right)^{0.66\pm0.37}
\label{eq:mbhwlq}
\end{multline}

The weaker dependence on \FWHM(\Hb) can be realized
when the BLR is elongated and parallel to the accretion disk.
There is accumulated evidence in the literature favoring a disk-like 
geometry for the BLR
\citep{Wills86, Laor06, Decarli08, Pancoast14, Shen14, Mejia18N,Wang19}. 
On the other hand, the BLR may be also dominated by outflows, which
are perpendicular to the line-of-sight. 
This scenario will favor a quasar reactivation idea. The outflow could rebuild 
the \Hb\ region \citep{Hryniewicz10}.
 
Systematic underestimation of \FWHM\ (and \mbhlit) may also be caused by a
strong influence of the \FeII\ pseudo-continuum in optics.  Such
phenomena is noticed by \citet{Plotkin15} for their sample of WLQs,
which have larger $R_{\rm opt,FeII}$ and narrower \Hb\ than most
reverberation mapped quasars. 

\citet{Mejia19} find that the dependence of \mbh\ on the observed
\FWHM\ of the Balmer lines for AGNs is close to linear rather (\mbh
$\propto$ \FWHM(\Hb)$^{0.82 \pm 0.11}$, when $f$ is a function of
the Full Width) than quadratic (\mbh $\propto$ \FWHM$^2$ with $f=$
const).  In our case, this relationship
for WLQs is a bit weaker (\mbh $\propto$ \FWHM(\Hb)$^{0.66 \pm
  0.37}$), but still compatible with the \citeauthor{Mejia19} result
within 1$\sigma$ error.  A similar or even the same behavior of
normal AGNs and WLQs suggests that both kind of sources have the same
dim nature of the velocity component and similar geometry of the BLR.
 
 The relationship between the widths of \MgII\ and  \Hb\ emission-lines is noticed in AGNs \citep[see][]{Shen08,Wang09}.
The estimation of BH masses based on those lines are consistent with each other.
However, a bias between the \CIV\ and \MgII\ mass estimation suggests that the \CIV\ estimator is severely affected 
by an outflow \citep{Baskin05, Trakhtenbrot12, Kratzer15}.  The authors argue that using \Hb\ or \MgII\ is better for BH mass estimation than \CIV.
Similar note is made by \citeauthor{Shen08}, whose results are based on 58 643 quasars from the SDSS catalog and who claim that 
the bias may be too large for individual objects using CIV estimator, but it is still consistent with \MgII\ and \Hb\ in the mean.
On the other hand, \citet{Wang09} have found that \FWHM(\MgII) is systematically smaller than \FWHM(\Hb) and the 
BH masses based on \MgII\ estimator show subtle deviations from those commonly used.
Referring to WLQs, \citet{Plotkin15} suggest that using \MgII\ line could cause bias to the mass measurements due to the big contribution of the \FeII\ pseudo-continuum. Thus, in this work, we base the estimations of BH masses on FWHM(\Hb) by trying to avoid \CIV\ and \MgII.

In our paper we fix the value of the radiative efficiency $\eta=0.18$.
However, we would like to check its influence on the results.
We again perform simulations and compare values of the best fits of BH masses and accretion rates in two cases, when $\eta=0.18$ and $\eta=0.36$. The accretion rates are 40-70\% higher and the BH masses are on average 20-30 \% less massive for higher~$\eta$ (for the first order approximation we have $\mmbh^2 \mdotm/\eta \simeq$ const). Please note that such a decline in BH masses cannot explain underestimation of \mbhlit\ and 
the choice of solutions with $\eta \nsim 0.18$ increases the $\chi^2$.

\begin{figure}
    \centering
    \includegraphics[width=0.480\textwidth]{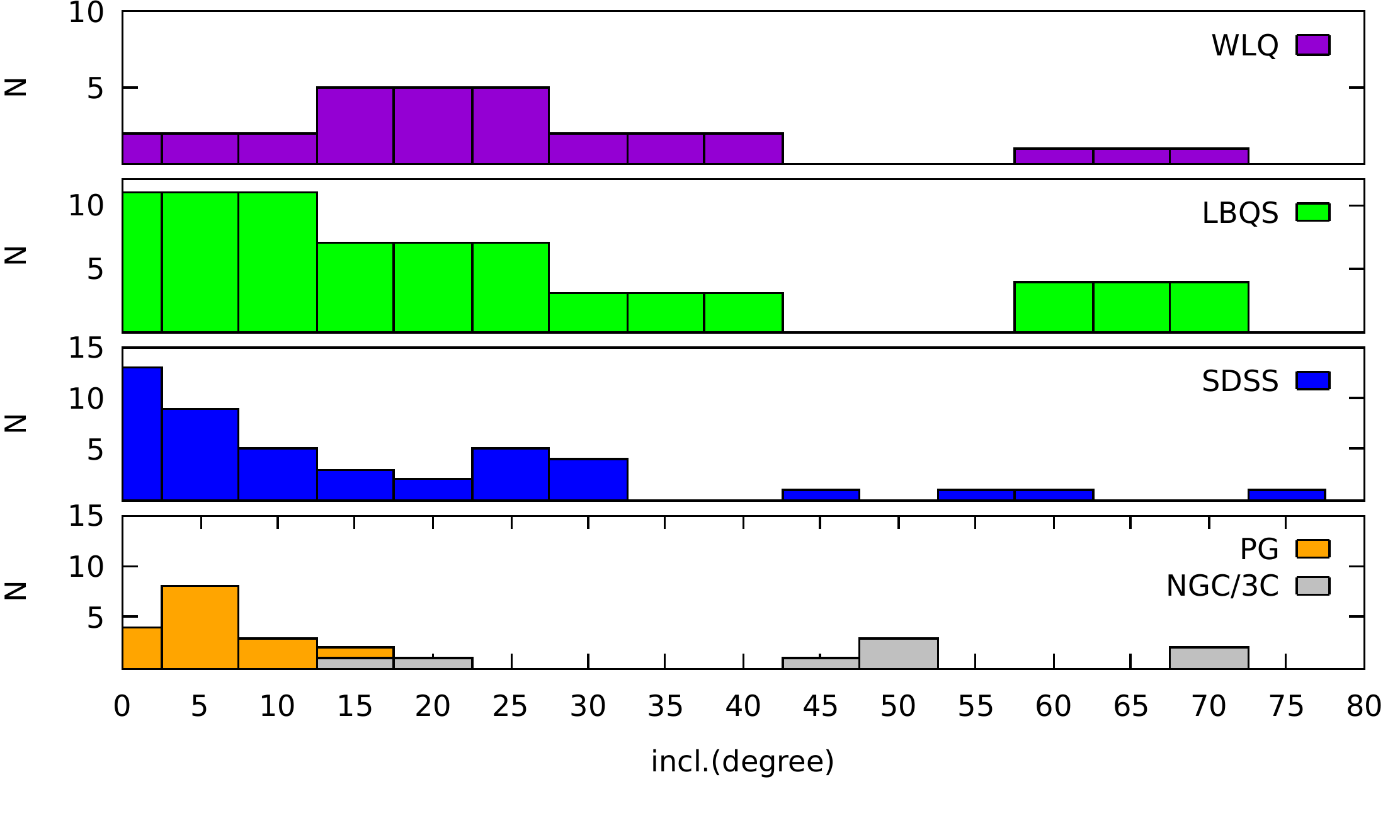}
    \caption{The distribution of inclinations in AGNs. Our sample of WLQs and LBQS quasars (the top panel and below).
    Blue histogram indicates SDSS quasars \citep{Wildy18}. The lowest panel shows PG quasars (orange histogram), 3C and Seyfert 1 galaxies (grey color). 
    The data come from \citet{Bian2002}.}
    \label{fig:incl_hist}
\end{figure}

The agreement for LBQS and discrepancy for WLQ quasars between their 
virial and the SED masses may suggest that \FWHM\ is a biased indicator 
of the virial velocity, due to the inclination of the emission region of 
the \Hb\ line. The BLR in WLQs could be less face-on than those in LBQS 
and other quasars. We compare inclinations of WLQs and LBQS quasars in 
our sample with those derived from SDSS quasars \citep{Wildy18}, PG quasars 
and Seyfert 1 galaxies \citep{Bian2002}. Generally, the main inclinations 
in LBQS are located at range 0-15$^{\circ}$, whereas WLQs are shifted toward higher values with the main peak at 15-30$^{\circ}$
(Fig. \ref{fig:incl_hist}). 
We point out the same average inclination values in SDSS and LBQS quasars.
It is worth to mention that fitted errors are significant. Thus, inclinations may be the same in all quasars. According to \citet{Collin04} (see their equation 11 and figure 8), bigger inclination in WLQs (e.g. $i_{LBQS} = 5^\circ \rightarrow i_{WLQ}=20^\circ$) together with the assumption of the flat BLR geometry ($H/\mRblr \lesssim 0.3$) cause underestimation of \FWHM\ by a factor $\gtrsim 2.5$. If we assume that the widths with values in the order of 2000 \kms are produced in the disk-like geometry, then BH masses should be heavier by 6-10 times. It is comparable with our analysis.
Higher widths, like those in LBQS quasars, could be formed in the BLR with spherical geometry. In this case the calculated BH masses do not require a correction.
It seems that the flatness of the BLR in the weak emission-line quasars plays a more important role than the inclination.
\section{Conclusion}
\label{sec:6}
In this work we have studied the accretion disk continua of 10
WLQs. The SMBH masses of those objects are estimated previously based
on the single-epoch virial BH mass method (\mbhlit). We create grid of 366000 models using
the Novikov-Thorne formulas. We adopt four parameters (\mbh, \dotm,
spin of BH and the line-to-sight inclination) to describe the observed
SED and compare obtained BH masses with those got from the literature.

Our main findings are:
\begin{enumerate}
    \item  Using the Novikov-Thorne model, we can describe very well the SED of WLQs. 
    
    \item The SMBH masses of WLQs, which are estimated based on
      \FWHM(\Hb), are underestimated.  On average, the masses are
      undervalued by 4-5 times. The median of this correction factor is 3.3.

    \item We propose the formula to estimate \mbh\ in WLQs based on
      their observed \FWHM(\Hb)\ and luminosities at
      5100\AA\ (Eq.~\ref{eq:mbhwlq}).  Our results suggest that
      selected WLQs have the accretion rates in the range $\sim$
      0.3-0.6.

    \item We support \citeauthor{Mejia19} result and confirm that the
      virial factor, $f$, depends on \FWHM. In this paper it $\propto$
      \FWHM(\Hb)$^{-1.34 \pm 0.37}$. The BLR is a non-spherical region.

    \item We suggest that WLQs are normal quasars in a reactivation stage, 
    in which the BLR region has the disk-like geometry.
    
\end{enumerate}

\section*{Acknowledgements}
We would like to thank the anonymous referee for useful comments that
improved the clarity of the paper. We also thank to Bo\.zena Czerny,
Ari Laor and Samuele Campitiello for helpful feedback and discussions.
The authors acknowledge support from the (Polish) National Science
Center Grant No.  2016/22/M/ST9/00583.  We would like to thanks
databases: The Sloan Digital Sky Survey (SDSS), the Two Micron All Sky
Survey (2MASS), Galaxy Evolution Explorer (GALEX) , the Wide-field
Infrared Survey Explorer (WISE) and VizieR service. This research has
made use of the NASA/IPAC Extragalactic Database (NED), which is
operated by the Jet Propulsion Laboratory, Caltech, under contract
with the NASA.
\medskip
\bibliography{apj}
\end{document}